\documentclass[useAMS,usenatbib,usegraphicx]{pasj01}
\usepackage{lscape}
\usepackage{color}
\usepackage{comment}
\usepackage{natbib, aas_macros}
\newcommand{\simgt}{\lower.5ex\hbox{$\; \buildrel > \over \sim \;$}}
\newcommand{\simlt}{\lower.5ex\hbox{$\; \buildrel < \over \sim \;$}}

\def\hMsol{\mathrel{h^{-1}M_\odot}}
\def\h70Msol{\mathrel{h_{70}^{-1}M_\odot}}
\begin{document} 
\Received{}%{yyyy/mm/dd}
\Accepted{}%{yyyy/mm/dd}
%\Published{yyyy/mm/dd}

\title{Multiwavelength study of X-ray Luminous Clusters
in the Hyper Suprime-Cam Subaru Strategic Program S16A field
\thanks{Based on data collected at Subaru Telescope, which is operated
by the National Astronomical Observatory of Japan.\\ Based on observations obtained with XMM-Newton, an ESA science mission with instruments and contributions directly funded by ESA Member States and NASA.}}

%%% begin:list of authors
% Do NOT capitalize all letters in "textsc".
\author{Keita \textsc{Miyaoka}\altaffilmark{1}}
\altaffiltext{1}{Department of Physical Science, Hiroshima University,
1-3-1 Kagamiyama, Higashi-Hiroshima, Hiroshima 739-8526, Japan}
\email{miyaoka@astro.hiroshima-u.ac.jp}

\author{Nobuhiro \textsc{Okabe}\altaffilmark{1,2,3}}
\altaffiltext{2}{Hiroshima Astrophysical Science Center, Hiroshima University, 1-3-1 Kagamiyama, Higashi-Hiroshima, Hiroshima 739-8526, Japan}
\altaffiltext{3}{Core Research for Energetic Universe, Hiroshima University, 1-3-1, Kagamiyama, Higashi-Hiroshima, Hiroshima 739-8526, Japan}

\email{okabe@hiroshima-u.ac.jp}

\author{Takao \textsc{Kitaguchi}\altaffilmark{3,1}}

\author{Masamune \textsc{Oguri}\altaffilmark{4,5,6}}
\altaffiltext{4}{Research Center for the Early Universe, University of Tokyo, Tokyo 113-0033, Japan}
\altaffiltext{5}{Department of Physics, University of Tokyo, Tokyo 113-0033, Japan}
\altaffiltext{6}{Kavli Institute for the Physics and Mathematics of the Universe (Kavli IPMU, WPI), University
of Tokyo, Chiba 277-8582, Japan}

\author{Yasushi \textsc{Fukazawa}\altaffilmark{1,2,3}}

\author{Rachel \textsc{Mandelbaum}\altaffilmark{7}}
\altaffiltext{7}{McWilliams Center for Cosmology, Department of Physics,
Carnegie Mellon University, 5000 Forbes Ave., Pittsburgh, PA 15213, USA}

\author{Elinor \textsc{Medezinski}\altaffilmark{8}}
\altaffiltext{8}{Department of Astrophysical Sciences, Princeton University, Princeton, NJ 08544, USA}

\author{Yasunori \textsc{Babazaki}\altaffilmark{9}}
\altaffiltext{9}{Department of Physics, Nagoya University, Aichi
464-8602, Japan}

\author{Atsushi J. \textsc{Nishizawa}\altaffilmark{10}}
\altaffiltext{10}{Institute for Advanced Research, Nagoya University Furocho, Chikusa-ku, Nagoya, 464-8602
Japan}
\author{Takashi \textsc{Hamana}\altaffilmark{11}}
\altaffiltext{11}{National Astronomical Observatory of Japan, Mitaka, Tokyo 181-8588, Japan}
\author{Yen-Ting \textsc{Lin}\altaffilmark{12}}
\altaffiltext{12}{Institute of Astronomy and Astrophysics, Academia
Sinica, P.O. Box 23-141, Taipei 10617, Taiwan}

\author{Hiroki \textsc{Akamatsu}\altaffilmark{13}}
\altaffiltext{13}{SRON Netherlands Institute for Space Research, Sorbonnelaan 2, 3584 CA Utrecht, The Netherlands}
\author{I-Non \textsc{Chiu}\altaffilmark{12}}

\author{Yutaka \textsc{Fujita}\altaffilmark{14}}
\altaffiltext{14}{Department of Earth and Space Science, Graduate School of Science, Osaka University, Toyonaka, Osaka 560-0043}
\author{Yuto \textsc{Ichinohe}\altaffilmark{15}}
\altaffiltext{15}{Department of Physics, Tokyo Metropolitan University,
1-1 Minami-Osawa, Hachioji, Tokyo 192-0397, Japan}
\author{Yutaka \textsc{Komiyama}\altaffilmark{11,16}}
\altaffiltext{16}{The Graduate University for Advanced Studies, 2-21-1 Osawa, Mitaka, Tokyo 181-8588, Japan}
\author{Toru \textsc{Sasaki}\altaffilmark{17}}
\altaffiltext{17}{Department of Physics, Tokyo University of Science, 1-3 Kagurazaka, Shinjyuku-ku, Tokyo 162-8601, Japan}
\author{Motokazu \textsc{Takizawa}\altaffilmark{18}}
\altaffiltext{18}{Department of Physics, Yamagata University, Kojirakawa-machi 1-4-12, Yamagata 990-8560, Japan}
\author{Shutaro \textsc{Ueda}\altaffilmark{19}}
\altaffiltext{19}{Institute of Space and Astronautical Science, Japan Aerospace Exploration Agency,3-1-1 Yoshinodai, Chuo-ku, Sagamihara, Kanagawa 229-8510, Japan}
\author{Keiichi \textsc{Umetsu}\altaffilmark{12}}

\author{Jean \textsc{Coupon}\altaffilmark{21}}
\altaffiltext{21}{Department of Astronomy, University of Geneva,
ch. d\'Ecogia 16, 1290 Versoix, Switzerland}
\author{Chiaki \textsc{Hikage}\altaffilmark{6}}
\author{Akio \textsc{Hoshino}\altaffilmark{22}}
\altaffiltext{22}{Department of Physics, Rikkyo University, Tokyo 171-8501, Japan}
\author{Alexie \textsc{Leauthaud}\altaffilmark{23}}
\altaffiltext{23}{Department of Astronomy and Astrophysics, University of California Santa
Cruz, Santa Cruz, CA 95064, USA}
\author{Kyoko \textsc{Matsushita}\altaffilmark{17}}
\author{Ikuyuki \textsc{Mitsuishi}\altaffilmark{9}}

\author{Hironao \textsc{Miyatake}\altaffilmark{24,6}}
\altaffiltext{24}{Jet Propulsion Laboratory, California Institute of Technology, Pasadena, CA 91109, USA}
\author{Satoshi \textsc{Miyazaki}\altaffilmark{11}}
\author{Surhud \textsc{More}\altaffilmark{6}}
\author{Kazuhiro \textsc{Nakazawa}\altaffilmark{25}}
\altaffiltext{25}{Department of Physics, The University of Tokyo, 7-3-1 Hongo, Bunkyo-ku, Tokyo 113-0033, Japan}
\author{Naomi \textsc{Ota}\altaffilmark{26}}
\altaffiltext{26}{Department of Physics, Nara Women's University, Kitauoyanishi-machi, Nara, Nara 630-8506, Japan}
\author{Kousuke \textsc{Sato}\altaffilmark{17,27}}
\altaffiltext{27}{Department of Physics, Saitama University,  255 Shimo-Okubo, Sakura-ku, Saitama, 338-8570}
\author{David \textsc{Spergel}\altaffilmark{8}}
\author{Takayuki \textsc{Tamura}\altaffilmark{19}}
\author{Masayuki \textsc{Tanaka}\altaffilmark{11}}
\author{Manobu M \textsc{Tanaka}\altaffilmark{28}}
\altaffiltext{28}{High Energy Accelerator Research Organization and The
Graduate University for Advanced Studies, Oho 1-1, Tsukuba, Ibaraki, Japan}
\author{Yousuke \textsc{Utsumi}\altaffilmark{2}}

%\altaffiltext{3}{C-Address of Institute}
%\email{ccccc@xxx.xxx.xx.xx}
%%% end:list of authors

%% `\KeyWords{}' always has to be placed before `\maketitle'.
 %Do NOT move this preamble from here!

\maketitle

\begin{abstract}
We present a joint X-ray, optical and weak-lensing analysis for X-ray
 luminous galaxy clusters selected from the MCXC (Meta-Catalog of X-Ray Detected Clusters of
Galaxies) cluster catalog 
in the Hyper Suprime-Cam Subaru Strategic Program (HSC-SSP) survey field
 with S16A data, As a pilot study of our planned series papers,
we measure hydrostatic equilibrium (H.E.) masses using {\it XMM-Newton} data
 for four clusters in the current coverage area out of a sample of 22 MCXC clusters. 
We additionally analyze a non-MCXC cluster associated with one MCXC cluster.
We show that H.E. masses for the MCXC clusters are correlated with cluster richness from
 the CAMIRA catalog \citep{Oguri17}, 
while that for the non-MCXC cluster deviates from the scaling relation.
The mass normalization of the relationship between the cluster richness
 and H.E. mass is compatible with one inferred by
 matching CAMIRA cluster abundance with a theoretical halo mass
 function.
The mean gas mass fraction based on H.E. masses for the MCXC clusters is $\langle
 f_{\rm gas} \rangle = 0.125\pm0.012$ at spherical overdensity $\Delta=500$, 
which is $\sim80-90$ percent of the cosmic mean baryon fraction,
 $\Omega_b/\Omega_m$, measured by cosmic microwave background
 experiments. 
We find that the mean baryon fraction estimated from X-ray and HSC-SSP
 optical data is comparable to $\Omega_b/\Omega_m$.
A weak-lensing shear catalog of background galaxies, combined with photometric
 redshifts, 
is currently available only for three clusters in our sample.
Hydrostatic equilibrium masses roughly agree with weak-lensing masses, albeit with large uncertainty.
This study demonstrates that further multiwavelength study
 for a large sample of clusters using X-ray,
 HSC-SSP optical and weak lensing data will enable us to understand
 cluster physics and utilize cluster-based cosmology.
\end{abstract}

\section{Introduction}

Galaxy clusters are the largest collapsed objects in the Universe,
and the evolution of the dark halo mass is sensitive to the growth
of matter density perturbations controlled by dark matter and dark energy.
Thus, observations of the high-mass exponential tail of the mass function over
wide redshift ranges can constrain cosmological parameters \citep[e.g.][]{Vikhlinin09b,Mantz16a}.

The anticipated wealth of data from both ongoing and upcoming
multiwavelength galaxy cluster surveys like Hyper Suprime-Cam Subaru Strategic Program
\citep[HSC-SSP;][]{Miyazaki12,Miyazaki15,HSC1styr,HSC1styrOverview}, Canada-France-Hawaii
Telescope Legacy Survey \citep[CFHTLS;][]{CFHTLS12}
the Dark Energy Survey \citep[DES;][]{DES16}, XXL \citep{XXL16}, 
Extended Roentgen Survey with an Imaging Telescope Array \citep[\emph{eROSITA};][]{eROSITA11},
\emph{Planck} \citep{Planck15CluterCosmology}, South Pole Telescope \citep[SPT;][]{SPTSZ15},
South Pole Telescope Polarimeter \citep[SPTPol;][]{SPTPol12},
 Atacama Cosmology Telescope \citep[ACT;][]{ACTSZ13} and Atacama
 Cosmology Telescope Polarimeter \citep[ACTPol;][]{ACTPOL16},
 now launches us into a new era of cluster-based cosmology and cluster study.
A persistent challenge that affects the ultimate scientific impact of
all of these surveys is the need for accurate measurements of the mass
for individual clusters.  

In the last two decades, X-ray observations
 \citep[e.g.][]{Vikhlinin06,Zhang08,Sun09,Martino14,Mahdavi13,Donahue14} of the intracluster medium
(ICM) have been used to measure gas temperature and density distributions and estimate
the total mass under the assumption of hydrostatic equilibrium
(H.E.). 
However, it is known that clusters are not exactly in H.E. because of
 some non-thermal phenomena in clusters such as radiative cooling and
 feedback from supernovae and active galactic nuclei (AGNs) in cluster
 central regions  \citep[e.g.][]{Kravtsov05,Pratt10,Planelles13}.
 Also the efficiency of accretion-shock heating of the infalling gas
 \citep[e.g.][]{Kawaharada10,Lapi10,Walker12b,Fujita13,Okabe14b,Avestruz16}
 is still not well understood. 
The deviation between the H.E. mass and an actual total mass depends on
 the hydrodynamical states of individual clusters.  
The mean deviation among a cluster sample is called ``mass bias''.
Indeed, the mass bias may be as one of the main causes of the tension
 in cosmological parameters obtained by the {\it Planck} cluster
 number counts \citep{Planck15CluterCosmology} and the {\it Planck} cosmic microwave
background (CMB) analysis \citep{Planck15Cosmology}. 
Therefore, X-ray observations have posed a challenge to this fundamental
 assumption.

On the other hand, weak lensing (WL) distortions of background galaxy images provide us with a
unique opportunity to reconstruct the mass distribution in clusters 
without any assumptions of dynamical states \citep{Bartelmann01},
making WL complementary to X-ray analysis.
In the past decade, a tremendous progress of WL analysis was made by prime focus
cameras at large ground-based telescopes, like Subaru/Suprime-Cam 
\citep[e.g.][]{Okabe08,Okabe10b,Okabe13,Okabe16,Oguri10b,Oguri12,Umetsu11,Umetsu16,Miyatake13,Medezinski15}
or wide field surveys \citep[e.g.][]{Mandelbaum06,Melchior16,Simet17}.
WL mass estimates are, however, sensitive to assumptions about the 3D shapes and halo orientations \citep[e.g.][]{Oguri05}
and substructures \citep[e.g.][]{Okabe14a} in the cluster gravitational potential, as well as any other large-scale structure between the lensed
sources and the observer \citep[e.g.][]{Hoekstra03}. Numerical simulations
\citep[e.g.][]{Meneghetti10,Becker11} have shown that 
WL mass estimates have scatter caused by a combination of the above effects.

In order to constrain the H.E. mass bias and to test the validity
of the H.E. assumption, which are of fundamental importance for
cosmological applications, previous studies
\citep[e.g.][]{Zhang10,vonderLinden14,Okabe14b,Donahue14,Hoekstra15,Smith16},
compiled a large number of clusters having both H.E. masses and WL
masses. 
They compared the two masses to indirectly constrain
the non-thermal pressure component involved in turbulence and/or bulk
motions and its radial dependence, 
assuming a random orientation of halo asphericity.
As before, joint studies based on
complementary X-ray, optical and WL datasets are definitely important in
the new era of cluster physics and cluster-based cosmology.

The Hyper Suprime-Cam Subaru Strategic Program \citep[HSC-SSP;][]{HSC1styr,HSC1styrOverview} is an ongoing wide-field imaging survey
using the HSC \citep{Miyazaki15} which is a new prime
focus camera of the 8.2m-aperture Subaru Telescope. 
The HSC-SSP survey is composed of three layers of different depths (Wide, Deep and UltraDeep). 
The Wide layer is designed to obtain five-band ($grizy$) imaging over $1400$~deg$^2$.
The HSC-SSP survey has both excellent imaging quality ($\sim$$0.''7$
seeing in $i$-band)
and deep observations ($r\simlt26$~AB~mag).  
The current status of the survey covers  $456$~deg$^2$ with non
full-depth and $178$~deg$^2$ with the full-depth and full-colour \citep{HSC1styr}. 
The HSC-SSP survey enables optical detection of two thousand galaxy clusters
\citep{Oguri17} in $\sim232$~deg$^2$ and will reconstruct mass distribution
of clusters up to $z\sim 1$ and beyond.

In this paper, we present H.E. mass measurements of galaxy clusters
in the current HSC-SSP field using {\it XMM-Newton} X-ray data, and compare X-ray observables with optical and WL
measurements. The H.E. mass measurement requires long
integration times with an X-ray satellite and therefore we selected X-ray luminous
galaxy clusters from an existing X-ray cluster catalog as a first study of the HSC-SSP survey.

The paper is organized as follows. We briefly summarize our target selection in Section
\ref{sec:HSCSSP}. The X-ray, optical and WL measurements are described
in Sections \ref{sec:xray}, \ref{sec:camira} and \ref{sec:MWL}, 
respectively. The main results and discussion are presented in Section
\ref{sec:result}. 
All results use  a flat $\Lambda$CDM cosmology with $H_0=70~{\rm 
km~s^{-1}Mpc^{-1}}$, $\Omega_{m,0}=0.3$ and $\Omega_\Lambda=0.7$.

\section{Target Selection} \label{sec:HSCSSP}

\begin{figure}
\begin{center}
\includegraphics[width=\hsize]{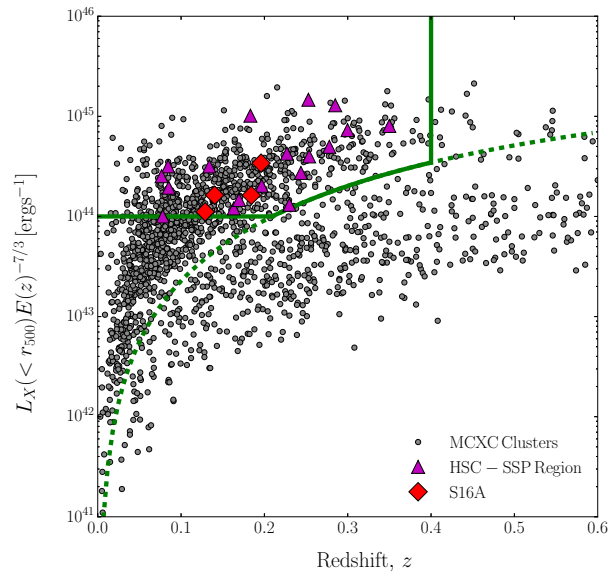}
\end{center}
\caption{Target selection: X-ray luminosity versus redshift for the MCXC clusters based on the {\it ROSAT} all sky survey. 
The green solid lines indicate our sample selection. The green dotted line
 is a flux threshold of $10^{-12}~{\rm erg~s^{-1}cm^{-2}}$.
Red diamonds and magenta triangles denote the targeting clusters in the
 paper and the full area ($\sim1400$ deg$^2$) of the HSC-SSP survey, respectively. }
\label{fig:target}
\end{figure}

With the aim of measuring H.E. masses, we select our sample of X-ray luminous clusters in the HSC-SSP field 
using the MCXC (Meta-Catalog of X-Ray Detected Clusters of
Galaxies) cluster catalog \citep{Piffaretti11} which is a homogeneously-measured cluster
catalog derived from several public catalogs based on the {\it ROSAT} all sky
survey. The cluster selection from the MCXC catalog satisfies the following criteria : $z<0.4$,
$L_X(<r_{500})E(z)^{-7/3}>10^{44}~{\rm erg~s^{-1}}$ and 
$f_X>10^{-12}~{\rm erg~s^{-1}cm^{-2}}$ in the HSC-SSP survey region, 
%and $>1\,{\rm count~s^{-1}}$ within $r_{500}$ in the HSC-SSP survey region, 
where $L_X$ is the X-ray luminosity in the $0.1-2.4$~keV energy band,
$f_X$ is the X-ray flux and
$E(z)=(\Omega_{m0}(1+z)^3+\Omega_\Lambda)^{1/2}$. 
Adopting the mass-luminosity scaling relation \citep{Piffaretti11},
the luminosity selection with the correction term $E(z)^{-7/3}$
may be assumed to be equivalent to the mass selection, $M_{500}\simgt 2\times 10^{14}\h70Msol$. 
Here, $M_{500}$ is the mass enclosed by the overdensity radius, $r_{500}$, 
inside of which the mean mass density is $500$ times the critical mass
density, $\rho_{\rm cr}$, at the redshift, $z$.
In the eventual full area of the HSC-SSP survey of $\sim1400$ deg$^2$, 
22 clusters can be selected from the MCXC all sky X-ray survey (Figure \ref{fig:target}). 
To date, four X-ray luminous clusters (Table \ref{tab:catalog}) are 
in an area suitable for investigating 
cluster physics with the HSC-SSP S16A data \citep[$\sim232$~deg$^2$;][]{Oguri17}.   
We obtained {\it XMM-Newton} data for the three clusters through our own
program (Table \ref{tab:catalog2}).
In the data analysis, we serendipitously observed a companion cluster to the west of 
MCXCJ1415.2-0030. As shown in Table \ref{tab:catalog}, its redshift is
very close to that of MCXCJ1415 and its richness, $N_{\rm cor}$, is
higher than the originally-selected cluster \citep{Oguri17}. 
We cannot rule out the possibility that the companion cluster is the dominant component of the system.  We additionally carry out X-ray analysis for this
cluster, because it is important to precisely measure a gas density
profile for MCXC1415 and the contamination of its companion cluster.
The cluster is hereafter referred to as MCXCJ1415.2-0030W. 
This paper compiles the analysis of the four MCXC clusters and the companion cluster in
the S16A field. The details of the X-ray analysis for the full sample will
appear in a future paper.

\begin{table*}
  \caption{Cluster sample: $^a$ Cluster name. $^b$ Alternative name. $^c$
 Cluster redshift from the MCXC catalog. $^d$ X-ray centroid. $^e$ BCG
 position. $^f$ The center of CAMIRA
 catalog. $^g$ Cluster optical richness. $^h$ WL mass measurement satisfying the
 the full-depth and full-colour conditions for the current HSC-SSP footprint.
 $^\dagger$ We retrieved the data from
 the CAMIRA catalog \citep{Oguri17}.
} \label{tab:catalog}

  \begin{center}
    \begin{tabular}{lccccccc}
      \hline
      \hline
Name$^a$ & Alternative name$^b$ & redshift$^c$ & $(\alpha,\delta)_{\rm
     xmm}^{d}$ & $(\alpha,\delta)_{\rm
     bcg}^{e\dagger}$ & $(\alpha,\delta)_{\rm
     camira}^{f\dagger}$ & $N_{\rm cor}^{g\dagger}$ & WL$^{h}$ \\
\hline
MCXCJ0157.4-0550 % amtokpc=1.38023722e+02
 & ABELL 281
 & 0.12890  
	     & (29.294, -5.869)
 & (29.279, -5.887)
 & (29.301, -5.918)
		     &  41.3
			 & no\\ 
MCXCJ0231.7-0451 & ABELL 362 
  & 0.18430  
 & (37.927, -4.882)
 & (37.922, -4.882) 
		 & (37.922, -4.883)
		     & 116.4
			 & yes\\ 
MCXCJ0201.7-0212 & ABELL 291 
  & 0.19600  
	     & (30.429, -2.196)
		 & (30.430, -2.197)
 & (30.445, -2.198) 
		     & 76.2
			 & no \\ 
MCXCJ1415.2-0030 &  ABELL 1882A & 0.14030  %1.48317937e+02
	     & (213.785, -0.491)
		 & (213.785, -0.494)
 & (213.785, -0.493) 
		     & 43.0
			 & yes \\ 
MCXCJ1415.2-0030W & ABELL 1882B  & 0.14400$^\dagger$  
	     & (213.601, -0.377)
		 & (213.600, -0.379)
 & (213.618, -0.330) 
		     &   68.8
			 & yes \\ 

     \hline
    \end{tabular}
  \end{center}
\end{table*}

\section{X-ray Analysis} \label{sec:xray}

In order to measure the total cluster mass with X-rays from the ICM gas, 
which is assumed to be in H.E. with the cluster gravitational potential, we need the gas density and temperature profiles.
European Photo Imaging Camera \citep[EPIC;][]{EPICMOS1,EPICPN} on
board the {\it XMM-Newton} satellite offers an
opportunity to perform extremely sensitive imaging/spectroscopic observations for clusters.
EPIC data  were analyzed with the ESAS (Extended Source Analysis Software) package \citep{Snowden2008}.
The details of the data analysis are described in the following sections.
In this work, we used SAS version 16.0.0 and HEAsoft version 6.19 with the latest CALDB as of November 2016.

\subsection{Data reduction}\label{sec:xdataR}
The EPIC data were processed and screened in the standard way by using the ESAS pipeline.
The data were filtered for intervals of high background due to soft
proton flares, defined to be periods when the rates were out of the $2\sigma$ range of a rate distribution.
Point sources are removed from three EPIC (MOS1, MOS2, and pn) images with simultaneous maximum likelihood PSF fitting.
The radius to mask a point source is chosen so that the surface brightness of the point source is one quarter of the surrounding background.
If the radius is less than half of the power diameter (HPD$\sim 15''$), we reset the radius to HPD.
Table \ref{tab:catalog2} summarizes the cluster data observed with {\it XMM-Newton}.

\begin{table*}
  \caption{X-ray data in the S16A field. $^a$Cluster name. $^b$Observational ids. $^c$
 Net exposure time of each instrument after the data
 reduction. $^\dagger$Data observed through our program. $^\natural$ Archival data.} \label{tab:catalog2}
  \begin{center}
    \begin{tabular}{lcccc}
      \hline
      \hline
Name$^a$ & obsid$^b$ & \multicolumn{3}{c}{Net exposure (ks)$^c$} \\
     &       & MOS1 & MOS2 & pn \\
\hline
MCXCJ0157.4-0550 
 & 0781200101$^\dagger$
 & 27.8
 &27.2
 &16.1 \\
MCXCJ0231.7-0451 
 & 0762870201$^\dagger$ 
 & 22.5 
&22.3
&15.5\\
MCXCJ0201.7-0212  
 & 0655343801$^\natural$ 
 & 22.4
&22.4
&14.9 \\
MCXCJ1415.2-0030  
 & 0762870501$^\dagger$
 & 19.3
 &19.0
  &13.0\\
MCXCJ1415.2-0030W 
 & 0145480101$^\natural$
 & 11.0 
 & 11.7 
 &7.0   \\ \hline
    \end{tabular}
  \end{center}
\end{table*}
\subsection{Spectral fit}

In order to determine the gas temperature profile, a spectral fit is
performed in the same way as in \cite{Snowden2008}, where all
spectra extracted from regions of interest 
are simultaneously fitted with a common model,
including particle and cosmic background components which are assumed to
be uniform across the detector except for instrumental lines. 
In this work, we used all three EPIC instruments of the MOS1,MOS2 and pn cameras.
Three spectra, one from each instrument, are extracted from concentric annuli centered on an
intensity-weighted centroid of the cluster.
As for the intensity-weighted centroid, we first select an
intensity peak and then iteratively determine intensity-weighted centroids.
within the radius of $500~h_{70}^{-1}$ kpc from the centroid.
At each iteration, we exclude regions,
of which sizes are the same as those of the excluded point sources
(Sec. \ref{sec:xdataR}), at their axially symmetric positions with
respect to the centroid computed by the previous iteration.
This process is important in order to avoid central shifts by the excluded point sources.
The calculation is converged within several iterations.
Each spectrum is binned in energy to have at least 35 counts per
spectral bin including background. 
Since the finite PSF effect can not be ignored for spectral
fits of cluster diffuse emission, we consider contaminations from
surrounding annuli using cross-talk auxiliary response files (ARFs) in
the spectral fitting.
The cross-talk contribution to the
spectrum in a given annulus from a surrounding annulus
 is handled as an additional model component.

The instrumental background spectrum,  which is stable with time,
 is modeled with data acquired with
 the filter wheel closed, available in ESAS CALDB, and is subtracted from the observed spectrum. 
The other particle backgrounds consisting of a continuum produced by soft protons and instrumental lines are determined by adding a power-law spectrum and narrow Gaussian lines with fixed central energies to the fitting model, respectively. 
The power-law model representing the soft proton background is added
only in MCXCJ0157.4-0550
%because, for the other clusters, high-rate events
%determined by the 2$\sigma$ cut of the rate distribution
%constitute only a few $\%$ of all the events and therefore are negligible.
because, for the other clusters,
the spectrum in the outermost region of interest is not affected by the contamination of 
the soft proton background and therefore corresponding model is negligible. 

The cosmic diffuse background consists of cosmic X-ray background (CXB), Galactic diffuse emission, and solar wind charge exchange (SWCX) emission lines, all of which are added to the fitting model. 
The CXB component is modeled with a power-law spectrum with a fixed index of 1.46 according to \cite{Snowden2008}.
The Galactic diffuse emission is fitted with the sum of absorbed and
non-absorbed thermal plasma emission models, with the temperature ranging from 0.25 to 0.7~keV and from 0.1 to 0.3~keV, respectively.
The SWCX lines are two narrow Gaussian models with fixed central energies of 0.56 and 0.65~keV, which correspond to OVII and OVIII lines, respectively.
The CXB and Galactic diffuse emission are constrained by simultaneously fitting a spectrum extracted from the $1^{\circ}$--$2^{\circ}$ annulus region surrounding the cluster using the ROSAT all sky survey (RASS) data \citep{RASS1997}.

The ICM emission spectrum is fitted by a thermal plasma emission model, APEC \citep{2001ApJ...556L..91S}, with the Galactic photoelectric absorption model, phabs \citep{1992ApJ...400..699B}.
In the identical annuli of the three EPIC detectors, each spectrum has common model parameters for ICM emission except for a normalization factor for cross-calibration. 
The metal abundance relative to solar from \cite{1989GeCoA..53..197A} in
each annulus is co-varied among the three instruments.
When the metallicity at large radii is not constrained,
it is the same as the value determined in the adjacent inner annulus.
%The power-law index and normalization for the soft proton background are common parameters among all the spectra and are allowed to vary.
The power-law indices of the soft proton background are
different free parameters in the MOS and pn, but are common in
all annuli for each detector.
The normalization for the soft proton background in individual annuli
varies according to a scale factor computed from ESAS
CALDB which contains actual
soft proton events \citep[see Fig.4 in ][]{Snowden2008}.
Cluster redshift, the hydrogen column density for the Galactic absorption, and instrumental line center energies are also common but fixed.
The hydrogen column densities use weighted averages from
\citet{2005A&A...440..775K} at cluster positions.

\begin{figure}
\begin{center}
   \includegraphics[width=60mm,angle=-90]{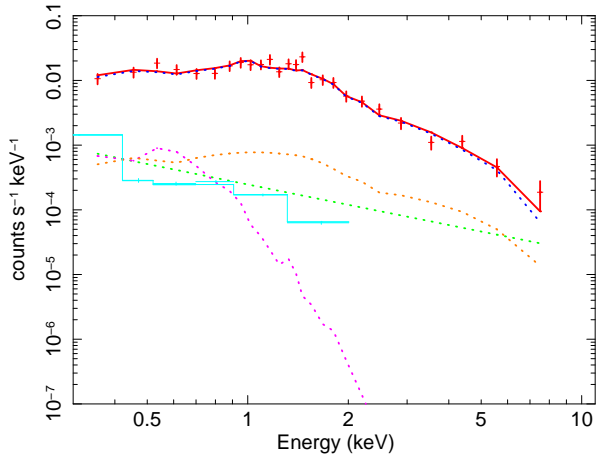}
\end{center}
\caption{Observed MOS1 spectrum in the center region of MCXCJ0157.4-0550
 with various fitting model components. In the actual fitting, all 24
 spectra are jointly  fitted with the common model. The red points show
 the observed spectrum with the instrumental background subtracted. The
 red solid line is the sum of all components. The blue, magenta, orange
 and green dotted lines are the thermal emission spectrum of the
 cluster, Galactic diffuse emission, CXB, and the residual soft proton
 component, respectively. The light blue line is the RASS spectrum.}\label{fig:spectra fit}
\end{figure}

To properly treat the finite size of the PSF in {\it XMM-Newton}
affect, we considered cross-talk ARFs \citep{Snowden2008} of neighboring
three or less annuli in the spectral fit. A change of best-fit
temperatures derived with and without cross-talk ARFs occurs at
very central region of which radial width is $\sim1'$. 
We summarize the result of the spectral-fit in Appendix \ref {sec:Appspecfit} and Table \ref{table:temp}.
Figure \ref{fig:spectra fit} is a typical spectrum of MOS1 in the center
region of  MCXCJ0157.4-0550. 
%For all 24 observed spectra in combination of 3 instruments with 8
%annuli, the value of $\chi^2$ is 1.304 with degrees of freedom of 111.
%For all 24 observed spectra in combination of 3 instruments with 8
%annuli, the number of degrees of freedom is 3868 and the number of free parameters is 111.

\subsection{Surface brightness profile}

The X-ray surface brightness profiles over entire detector regions are derived from the 0.4--2.3~keV image, from which the instrumental background is already subtracted.
We assume that the surface brightness profile follows a linear combination of $\beta$ models \citep{Cavaliere76,Jones84}, to analytically describe multi-scale and/or multi-component of the X-ray emitting gas. The model profile is described as follows,
\begin{eqnarray}
 S_X^{\rm tot}(R)=\sum_{i=1}^n S_{X,i}(R) + B 
\end{eqnarray}
where $S_{X,i}$ is a single $\beta$ model,
\begin{eqnarray}
 S_X(R)=S_0 \left (1+\left(R/r_{c}\right)^2\right)^{1/2-3\beta},  \label{eq:beta}
\end{eqnarray}
and $R$ is the projected distance from the center and $B$ is a
constant offset representing the CXB of each instrument. 
We monitor whether the best-fit $B$s are in good agreement with the blank sky. 
We confirmed that the surface brightness in the target fields and
blank sky are fairly constant at radii where the CXB dominates.
The $\beta$ model is convolved with each corresponding instrument 
point spread function (PSF) and is simultaneously fitted to the corresponding surface 
brightness profile of all three EPIC detectors; $\chi^2=\chi^2_{\rm
MOS1}+\chi^2_{\rm MOS2}+\chi^2_{\rm pn}$. 
We first fit the $\beta$ model to the data and, if it yields an apparent
poor fit,
we add another component to the model. All our data can be well
expressed by $n\le 2$. Given the best-fit parameters,
we can decompose the three dimensional density profile with a linear combination,
\begin{eqnarray}
 n_i^2(r)=n_{0,i}^2\left(1+\left(r/r_{c,i}\right)^2\right)^{-3\beta_i}.
\end{eqnarray}
Here, $r$ is the three-dimensional distance from the center.
We compute the emissivity in the given energy band from the best-fit parameters of spectral analysis 
and conversion factors between $S_X$ and $n_e$ considering detector sensitivities. 
The surface brightness is in general measured over larger radii than
that for spectral analysis, and the conversion factor at large radii is extrapolated with
a linear function of the radius. If we convert using the central
emissivity only, 
the electron number density is underestimated by $\sim10\%$ at $\sim r_{500}$. 

To fairly model multiple components for an internal substructure in
MCXCJ0157.4-0550 (Section \ref{subsec:MCXCJ0157}; Figures \ref{fig:sx_mcxcj0157} and \ref{fig:mcxcj0157})
 and a contamination from the close-pair cluster,
MCXCJ1415.2-0030 and MCXCJ1415.2-0030W (Section \ref{subsec:MCXCJ1415}), 
we take into account X-ray emission from gas components offset from
the main cluster when determining the surface brightness.
The off-centering effect in the surface brightness is calculated
by
\begin{eqnarray}
 S_{\rm X,off}(R)=\frac{1}{2\pi}\int_0^{2\pi} d\theta S_{\rm X}(\sqrt{R^2+d_{\rm                           
  off}^2-2Rd_{\rm off}\cos\theta}). \label{eq:Sxoff}
\end{eqnarray}
Here $d_{\rm off}$ is the off-centering distance on the sky.
When we model without the off-centering effect, the outer slope $\beta$
is misestimated and the H.E. mass estimates are biased.
As a sanity check, we compute the surface brightness profiles excluding
the area within which the best-fit number density for the
off-centered component is more than $10^{-2}$ of its central density and confirm that the best-fit results do not change.

\subsection{Temperature profile}
The temperature profile is modeled with a generalized universal profile \citep{Martino14,Okabe14b},
\begin{eqnarray}
 T_{\rm 3D}(r)=T_0 \frac{(r/r_t)^a}{(1+(r/r_t)^2)^{c/2}}.
\end{eqnarray}
The temperature profile projected along the line-of-sight is estimated
with a weight $w$,
\begin{eqnarray}
 T_{\rm 2D}=\frac{\int T_{\rm 3D} w dV }{\int w dV},
\end{eqnarray}
in each annulus.
We here assume the spectroscopic-like temperature \citep{Mazzotta04,Martino14} with $w=n^2 T_{3D}^{-3/4}$.
When the low photon-statistics prevents us from constraining a temperature
profile, we assume the inner slope $a=0$ and/or the outer slope $c=1$, following an universal temperature profile
out to $r_{200}$ based on joint Subaru WL and {\it Suzaku} X-ray analysis
\citep{Okabe14b}. 
We also assume a constant profile for MCXCJ1415.2-0030W because the
temperatures are measured only in two annuli.
The measurement uncertainty for the number density is also propagated to
the temperature fitting.

\subsection{H.E. Mass profile}\label{subsec:MHE}
%In the previous subsection, the derivation of three-dimensional density and temperature profiles 
%based on a forward modeling method \citep[e.g.][]{Martino14,Meneghetti10,Vikhlinin06} shows.  
%Given the best-fit models, we calculate gas mass
%and the total gravitating mass with H.E. assumption. 
%Measurement uncertainties for all quantities are estimated by a covariance
%error matrix and properly propagated to the gas and total masses.

Given the best-fit parameters, the three-dimensional spherical total mass is estimated
with the H.E. assumption, 
\begin{eqnarray}
M_{\rm H.E.}(r)=-\frac{k_B T_{\rm 3D}(r) r}{\mu m_p G}\left[\frac{d\ln \rho_g(r)}{d \ln r}+ \frac{d\ln
		      T_{\rm 3D}(r)}{d \ln r}\right],
\end{eqnarray}
where $\mu=0.5964$ is the mean molecular weight for the metallicity
$Z=0.3$ and $\rho_g =1.9257 \mu n_e m_p$ is the gas density
profile. Here, $n_e$ is the electron density and $m_p$ is the proton mass.
We then estimated the total mass, $M_{500}$, within $r_{500}$ and 
the spherical gas mass, $M_{\rm gas}(r)$, calculated by integrating
$\rho_g$ out to the radius, $r_{500}$.

\begin{table*}
  \caption{H.E. mass, WL mass, gas mass and stellar mass estimations at
 $\Delta=500$. Gas and stellar masses are enclosed within $r_{500}$ derived
 from H.E. masses. } \label{tab:mass}
  \begin{center}
    \begin{tabular}{lcccc}
      \hline
      \hline
     Name & %H.E. mass & WL mass & Gas mass & Stellar mass \\
      $M_{500}^{\rm H.E.}$  & $M_{500}^{\rm WL}$ & $M_{\rm gas}(<r_{500}^{\rm H.E.})$ & $M_{*}(<r_{500}^{\rm H.E.})$  \\
    & ($10^{14}\h70Msol$) & ($10^{14}\h70Msol$) & ($10^{13}h_{70}^{-5/2}M_\odot$) & ($10^{12}h_{70}^{-2}M_\odot$) \\
     \hline
MCXCJ0157.4-0550 &
$1.37_{-0.08}^{+0.09}$
	 & -
	     & $2.12_{-0.11}^{+0.12}$
     & $2.63_{-0.05}^{+0.05}$ \\
     MCXCJ0231.7-0451&
	 $3.43_{-0.65}^{+0.77}$
	 &  $7.96_{-1.89}^{+2.58}$
	     & $4.39_{-0.40}^{+0.42}$
  & $8.87_{-0.34}^{+0.31}$ \\
MCXCJ0201.7-0212& 
	 $3.21_{-0.44}^{+0.51}$
	 & -
	     & $4.28_{-0.30}^{+0.32}$
		 &$5.75_{-0.16}^{+0.15}$ \\
   MCXCJ1415.2-0030 &
$1.54_{-0.23}^{+0.34}$
	 & $2.09_{-0.90}^{+1.43}$
	 & $1.08_{-0.12}^{+0.16}$
         & $5.39_{-0.18}^{+0.23}$ \\
MCXCJ1415.2-0030W & $0.44^{+0.07}_{-0.07}$
	 & $0.80_{-0.58}^{+0.87}$
	     & $0.60_{-0.15}^{+0.17}$
 & $3.78_{-0.44}^{+0.37}$ \\
     \hline
    \end{tabular}
  \end{center}
\end{table*}

\section{Optical Catalog} \label{sec:camira}

We retrieved the cluster richness, $N_{\rm cor}$, and the stellar masses, $M_*$, from
the CAMIRA cluster catalog \citep{Oguri17} which is constructed using the HSC-SSP Wide S16A data.
The CAMIRA algorithm makes use of a stellar population synthesis model
to predict colours of red sequence galaxies at a given redshift for an
arbitrary set of bandpass filters and a three dimensional richness map with a compensated spatial filter.
The details of the CAMIRA cluster algorithm are described in \cite{Oguri14b} and \cite{Oguri17}. 
The smoothing scale for the compensated spatial filter is
$R_0=0.8h^{-1}{\rm Mpc}$ in physical units.
The total stellar mass for red galaxies of each cluster is estimated by
convolution with the spatial filter.
Since blue galaxies are a subdominant component in stellar mass at
low redshifts, we estimate the stellar mass only using red galaxies.  
We confirmed that the photometric redshifts provided by the CAMIRA
cluster catalog excellently agree with the spectroscopic redshifts.

We here use the stellar masses rather than optical luminosities
because the HSC-SSP multi-band datasets are capable of estimating the
stellar masses \citep{Oguri17}.
Following \cite{Miyazaki15}, 
we convert the total stellar masses ($M_*^{\rm CAMIRA}$) of each CAMIRA cluster into one
enclosed within a measurement radius ($M_*(<r_\Delta)$) .
%the stellar masses within a measurement radius from a total
%stellar mass of each CAMIRA cluster \citep{Oguri17} because the CAMIRA
%catalog provides us with the total stellar mass. 
The conversion factor, $A\equiv M_*(<r_\Delta)/M_*^{\rm CAMIRA}$, can be
calculated as follows. 
We first assume that the stellar mass density profile is described by
a universal mass density profile \citep[][hereafter NFW]{NFW96,NFW97} 
with the mass and concentration relation of \cite{Diemer14} with the
Planck cosmology \citep{Planck15Cosmology}.
Here, the NFW profile is expressed in the form:
\begin{equation}
\rho_{\rm NFW}(r)=\frac{\rho_s}{(r/r_s)(1+r/r_s)^2},
\label{eq:rho_nfw}
\end{equation}
where $\rho_s$ is the central density parameter and $r_s$ is the scale
radius. The halo concentration is defined by $c_\Delta=r_\Delta/r_s$,
where $r_\Delta$ is the overdensity radius.
Given the mass and its assumed concentration,
%$A=\bar{\Sigma}_{*}/\bar{\Sigma}_{*,\infty}$ is  
the conversion factor, $A$, is obtained as 
the ratio derived by integrating the projected NFW profile out to the measurement radius and
to infinity,  with a convolution of the spatial filter. 
When a measurement center is offset from the CAMIRA center, 
we calculate the azimuthally averaged, projected NFW density around the
measurement center and then integrate it out to the measurement radius
in a similar manner to the equation \ref{eq:Sxoff}.
As mentioned in \cite{Miyazaki15}, 
this correction technique takes into account the three dimensional deprojection. 
We measure stellar masses at $\Delta=500$
derived by X-ray mass measurements and estimate gas and baryon
fractions with the X-ray gas measurements (Section \ref{subsec:fb}).
Some very luminous galaxies in clusters are missed by flags in the CAMIRA catalog 
because their luminosity cannot be accurately measured due to saturation. 
If an offset between the CAMIRA and X-ray centers are large, 
the number of missing luminous galaxies is relatively large.
To include missing luminous galaxies, 
we add stellar masses of luminous member galaxies identified by SDSS
spectroscopic observations. The stellar masses and gas fractions including the luminous
galaxies are $\sim 3-47\%$ and $\simlt 14\%$ higher, respectively.

\section{Weak-lensing Mass Measurement} \label{sec:MWL}

We carry out WL analysis using the WL shear data estimated by 
re-Gaussianization method \citep{Hirata03} which is implemented in
the HSC pipeline \citep[see details in][]{HSCWL1styr}. 
Both precise shape measurement and photometric redshift estimation are essential
for the WL related studies, which offers the strict
conditions on the depth of data and the availability of five-band photometry.
For this reason, the WL shear catalog is restricted to the full-depth and
full-colour footprint. 
Three out of five clusters are located in the those regions, namely,
MCXCJ0231.7-0451, MCXCJ1415.2-0030 and MCXCJ1415.2-0030W, for which we
measure WL masses (Table \ref{fig:target}).

In cluster lensing, a contamination of unlensed member galaxies in
the source catalog significantly underestimates lensing signals at small
radii, because the fraction of member and background galaxies is increasing
with decreasing radius. 
It is known as the dilution effect \citep{Broadhurst05}.
Previous studies \citep[e.g.][]{Okabe13,Okabe14b,Okabe16b} have shown that if there is no background selection, 
lensing signals for massive clusters at $z\sim0.2$ are underestimated by $\sim40\%$. 
Previous studies \citep{Okabe13,Okabe16b,Medezinski10,Medezinski15,Umetsu16} 
securely selected background galaxies using colour information 
and succeeded in keeping the level of contamination below a few percent. 
We here briefly summarize the method. 
It is very difficult to discriminate between faint members and
background galaxies by magnitude information because of large
photometric uncertainty and the intrinsic scatter of color distribution.
We select a colour space region in which member galaxies
are negligible by monitoring a consistency among three independent
information of colour, lensing signal and available, external photometric redshift catalog \citep{Ilbert13}.
Since passively-evolving member galaxies are localized in colour
space, the mean tangential distortion strength in the colour space close
to the red sequence is significantly underestimated 
because member galaxies are not lensed.
 However, the mean lensing signals, which are computed by the ensemble
 shear and the photometric redshift, outside of red-sequence color-space
 are flatten due to a reduction in contamination by member galaxies. 
By modeling the colour distribution of faint member galaxies, we have succeed
in keeping the contamination limit at less than a few $\%$. In this
procedure, we considered both shape noises and errors of photometric redshifts.

Based on a similar philosophy, \cite{Medezinski17} have developed a new
scheme to make a secure selection of background galaxies. 
We utilized lensing signals and four-band magnitudes ($griz$) of the HSC-SSP survey, internal
photometric redshifts \citep{HSCPhotoz17} computed by machine learning
\citep[MLZ;][]{MLZ14} calibrated with spectroscopic data.
We have succeeded in selecting background galaxies in the $rz$ and $gi$ colour plane 
as the best combination for the HSC-SSP survey.
Based on \citet[see for details]{Medezinski17}, 
we select background galaxies for WL mass measurements \citep[see
also][]{Miyatake17}.
The number of background galaxies after the color cuts is $\simlt 11\,{\rm
arcmin}^{-2}$.

Given the shape catalog of background galaxies,
we measure the reduced tangential shear $\langle \Delta \Sigma_{+}\rangle(r_i)$
computed by azimuthally averaging the measured galaxy ellipticity,
$e_\alpha$ ($\alpha=1,2$), for $n$-th galaxy in a given
$i$-th annulus ($r_{\rm
inn}<r_i<r_{\rm out})$, centering at the
brightest cluster galaxies (BCGs),
\begin{eqnarray}
\langle \Delta \Sigma_{+}\rangle(r_i)= \frac{\sum_n e_{+,n} w_{n} \langle \Sigma_{{\rm
 cr},n}^{-1}\rangle^{-1}}{2 {\mathcal R}_i (1+K_i) \sum_n w_{n}}, \label{eq:g+}
\end{eqnarray}
where the tangential ellipticity is
\begin{eqnarray}
e_{+}&=&-(e_{1}\cos2\varphi+e_{2}\sin2\varphi),
\label{eq:gt}
\end{eqnarray}
and the inverse of the mean
critical surface mass density ($\langle \Sigma_{{\rm cr}}^{-1}\rangle$),
the weighting function ($w_n$), the shear responsivity
(${\mathcal R}$), and the calibration factor ($K$) are defined below.
The inverse of the mean critical surface mass density is computed by
the probability function $P(z)$ of the MLZ photometric redshift, 
\begin{eqnarray}
 \langle \Sigma_{{\rm cr}}^{-1}\rangle =
  \frac{\int^\infty_{z_l}\hat{\Sigma}_{{\rm
  cr}}^{-1}(z_l,z_s)P(z_s)dz_s}{\int^\infty_{0}P(z_s)dz_s}. \label{eq:Sigma_cr_inv}
\end{eqnarray}
Here, $z_l$ and $z_s$ are the cluster and source redshift, respectively.
The critical surface mass density for individual background galaxies is expressed by $\hat{\Sigma}_{{\rm cr}}=
c^2D_s/4\pi G D_l D_{ls}$, where $D_s$, $D_l$ and $D_{ls}$ are the
angular diameter distances from the observer to the sources,  from the observer to the lens,  
and from the lens to the sources, respectively.
Since $\hat{\Sigma}_{\rm cr}$ becomes zero for $z_s<z_l$, the lower
bound of the integration is truncated by $z_l$.
The weighting function $w_n$ is given by
\begin{eqnarray}
w_n=\frac{1}{e_{{\rm rms},n}^2+\sigma_{e,n}^2}\langle \Sigma_{{\rm cr},n}^{-1}\rangle^2\label{eq:weight},
\end{eqnarray}
where $e_{\rm rms}$ and $\sigma_e$ are the root mean square of
intrinsic ellipticity and the measurement error per component
($\alpha=1$ or $2$), respectively. Here, the intrinsic ellipticity
expresses the ellipticity for the intrinsic shape of galaxies.
The shear responsivity is computed based on the ellipticity definition,
\begin{eqnarray}
{\mathcal R_i}= 1-\frac{\sum_n e_{\rm rms,n}^2 w_n}{\sum_n w_n},
\end{eqnarray}
 \citep[see also][]{Mandelbaum05,Reyes12}. 
We correct the measured values using the shear calibration factor
$(m,c)$ for individual objects \citep{HSCWL1styr}, 
because of systematic error of shape measurements. 
The measured ensemble shear, $\langle \gamma \rangle$, can be
expressed by $(1+m) \gamma_{\rm true}+c$ with the input shear
$\gamma_{\rm true}$, as defined by STEP (Shear TEsting Programme) simulations
\citep{Heymans06,Massey07}. Here, 
a multiplicative calibration bias $m$ and an additive residual shear
offset $c$ are estimated based on GREAT3-like simulations
\citep{HSCGREAT03, Mandelbaum14,Mandelbaum15}  as a part of the
GREAT (GRavitational lEnsing Accuracy Testing) project. 
The calibration factor, $K$, is computed by
\begin{eqnarray}
 K_i=\frac{\sum_n m_n w_n}{\sum_n w_n}.
\end{eqnarray}
We also conservatively subtract  
\begin{eqnarray}
\tilde{c}_i=\frac{\sum_n c_{+,n} w_{n} \langle \Sigma_{{\rm
 cr},n}^{-1}\rangle^{-1}}{(1+K) \sum_n w_{n}}
\end{eqnarray}
from $\langle \Delta \Sigma_{+}\rangle(r_i)$  (equation
\ref{eq:g+}). The additional offset term is negligible ${\mathcal
O}(<10^{-4})$ compared to $\langle \Delta
\Sigma_{+}\rangle\sim{\mathcal O}(10^{-1})$.

Following \cite{Okabe16b}, we employ a maximum-likelihood method to model the shear profiles, and
express the log-likelihood as follows:
\begin{eqnarray}
-2\ln {\mathcal L}&=&\ln(\det(C_{ij})) +  \label{eq:likelihood} \\
 &&\sum_{i,j}(\Delta \Sigma_{+,i} - f_{{\rm model}}(r_i))C_{ij}^{-1} (\Delta
 \Sigma_{+,j} - f_{{\rm model}}(r_j)), \nonumber
\end{eqnarray}
where the subscripts $i$ and $j$ are for the $i-$ and $j-$th radial bins, respectively.
We adopt that $r_i$ is the weighted harmonic mean radius of the
background galaxies. 
Here, $f_{\rm model}$ is the reduced shear prediction for a specific
mass model,
\begin{eqnarray}
 f_{{\rm model}}(r_i)=\frac{\Delta \tilde{\Sigma}_{\rm model}(r_i)}{1 - {\mathcal L}_{z,i}
  \Sigma_{\rm model}(r_i)) }, \label{eq:g+model}
\end{eqnarray}
with $\Delta \tilde{\Sigma}_{\rm model}=\gamma_+\langle \Sigma_{\rm
cr,n}^{-1}\rangle^{-1}$ and $\Sigma_{\rm model}=\kappa \langle \Sigma_{\rm cr,n}^{-1}\rangle^{-1}$.
Here, $\gamma_+$ is the dimensionless tangential shear and $\kappa$ is
the convergence, respectively.
The covariance matrix, $C=C_g+C_s+C_{\rm LSS}$, in equation~\ref{eq:likelihood} is composed
of the shape noise $C_{g}$, the uncertainty of source redshift $C_s$ and the photometric redshift error computed by
$P(z)$ 
and uncorrelated large-scale structure (LSS), $C_{\rm LSS}$, along the
line-of-sight \citep{Schneider98}.
The covariance matrix for shape noise is obtained as weighted
variance 
\begin{eqnarray}
C_{g,ij}=\left(\frac{1}{4 R_i^2 (1+K_i)^2 \sum w_n}\right) \delta_{ij},
\end{eqnarray}
where $\delta_{ij}$ is a kronecker delta function.
The photometric error for individual galaxies is estimated by
\begin{eqnarray}
 \delta \langle \Sigma_{{\rm cr}}^{-1}\rangle =
  \frac{\int^\infty_{z_l}(\hat{\Sigma}_{{\rm cr}}^{-1}(z_l,z_s)-\langle \Sigma_{{\rm cr}}^{-1}\rangle)^2P(z_s)dz_s}{\int^\infty_{0}P(z_s)dz_s},
\end{eqnarray}
and then we propagate it into the measurement as $C_s$.
The covariance matrix, $C_{\rm LSS}$, is given by
\begin{eqnarray}
 C_{{\rm LSS},ij}= \langle \Sigma_{{\rm cr},i}^{-1}\rangle^{-1} \langle \Sigma_{{\rm cr},j}^{-1}\rangle^{-1} \int \frac{ldl}{2\pi}P_{\kappa}(l) J_2(l\theta_i) J_2(l\theta_j),
\end{eqnarray}
where $P_\kappa(l)$ is the weak-lensing power spectrum
\citep[e.g.][]{Schneider98,Hoekstra03}, calculated by multipole $l$,
the source redshift, and a given cosmology. We employ the Planck
cosmology \citep{Planck15Cosmology} for $\Omega_{b0}$, $\sigma_8$ and the spectral index $n_s$.
Here, $J_2(l\theta_i)$ is the Bessel function of the first kind and
second order at the $i$-th annulus \citep{Hoekstra03}.

The source redshift at each radial bin are calculated from
lensing-efficiency weighted value, as follows
\begin{eqnarray}
 {\mathcal L}_{z,i} = \frac{\sum_n \langle \Sigma_{\rm cr,n}^{-1}\rangle w_n
  }{\sum_n w_n}.
\end{eqnarray}

Following numerical simulations \citep{NFW96} and previous observational
results \citep[e.g.][]{Okabe13,Okabe16,Oguri12,Umetsu16}, we adopt the
NFW profile \citep{NFW96} as the mass model. 
Similar to the X-ray analysis, we define the three-dimensional spherical
mass, $M_{500}$, enclosed by the radius, $r_{500}$. We basically fit for two
parameters : the mass and the halo concentration.
The resulting masses are shown in Table \ref{tab:mass}.

\section{Results and Discussion} \label{sec:result}

We carried out X-ray analysis and joint X-ray and optical analysis
for the four MCXC clusters and the non-MCXC cluster (Table \ref{tab:catalog}) in the current coverage region
for the HSC-SSP survey. As mentioned in Section \ref{sec:MWL},
the sample for WL mass measurement is only the three clusters (Table \ref{tab:catalog}) due to
the full-colour and full-depth conditions. %Thus, the full datasets are not yet available. 
Thus, we shall use the WL mass estimates only for X-ray and WL mass comparisons.

Since this paper is a sort of pilot studies to directly compare
multi-wavelength datasets, we shall first discuss results for individual
clusters based on both X-ray and optical datasets.
%It is well-known that central positions derived by optical finders are
%offset from X-ray ones, of which details are discussed in \citet{Oguri17}.
%It would be caused by algorithm to use sparse distribution of galaxies,
%physical reasons or these combinations.
%We also shall show individual cluster case to obtain clues of the offset.}
We then perform studies of a correlation between H.E. masses and
richness, a mass comparison and baryon fraction for the current sample of clusters.

\subsection{MCXCJ0157.4-0550 } \label{subsec:MCXCJ0157}

\begin{figure}
\begin{center}
   \includegraphics[width=\hsize]{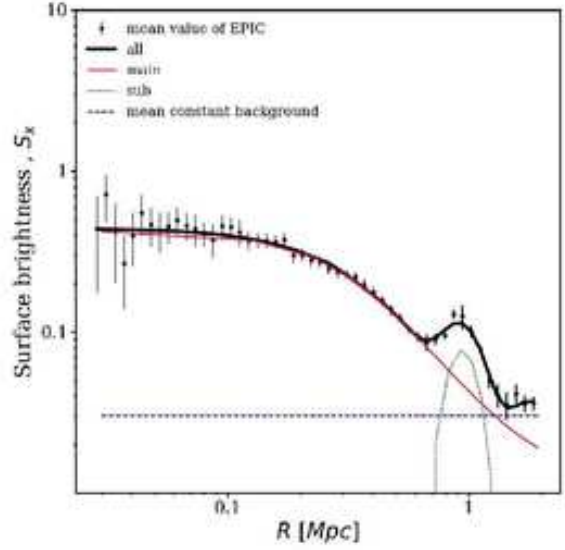}
\end{center}
\caption{The X-ray surface brightness profile for MCXCJ0157.4-0550 in arbitrary units
 averaged over three instruments. The projected distance is described by $R$.
The profile is corrected with different effective areas and background levels of MOS and pn. 
The center is the flux-weighted X-ray centroid of the western main cluster. 
Thin red solid, green dotted and blue dashed lines show the model
 profile of the western main cluster, the off-centered sub cluster and the
 mean constant background, respectively. The thick black solid line is the
 sum over all components, which describes the observed profile very well.
}\label{fig:sx_mcxcj0157}
\end{figure}

\begin{figure*}
\begin{minipage}{0.5\hsize}
\begin{center}
   \includegraphics[width=\hsize]{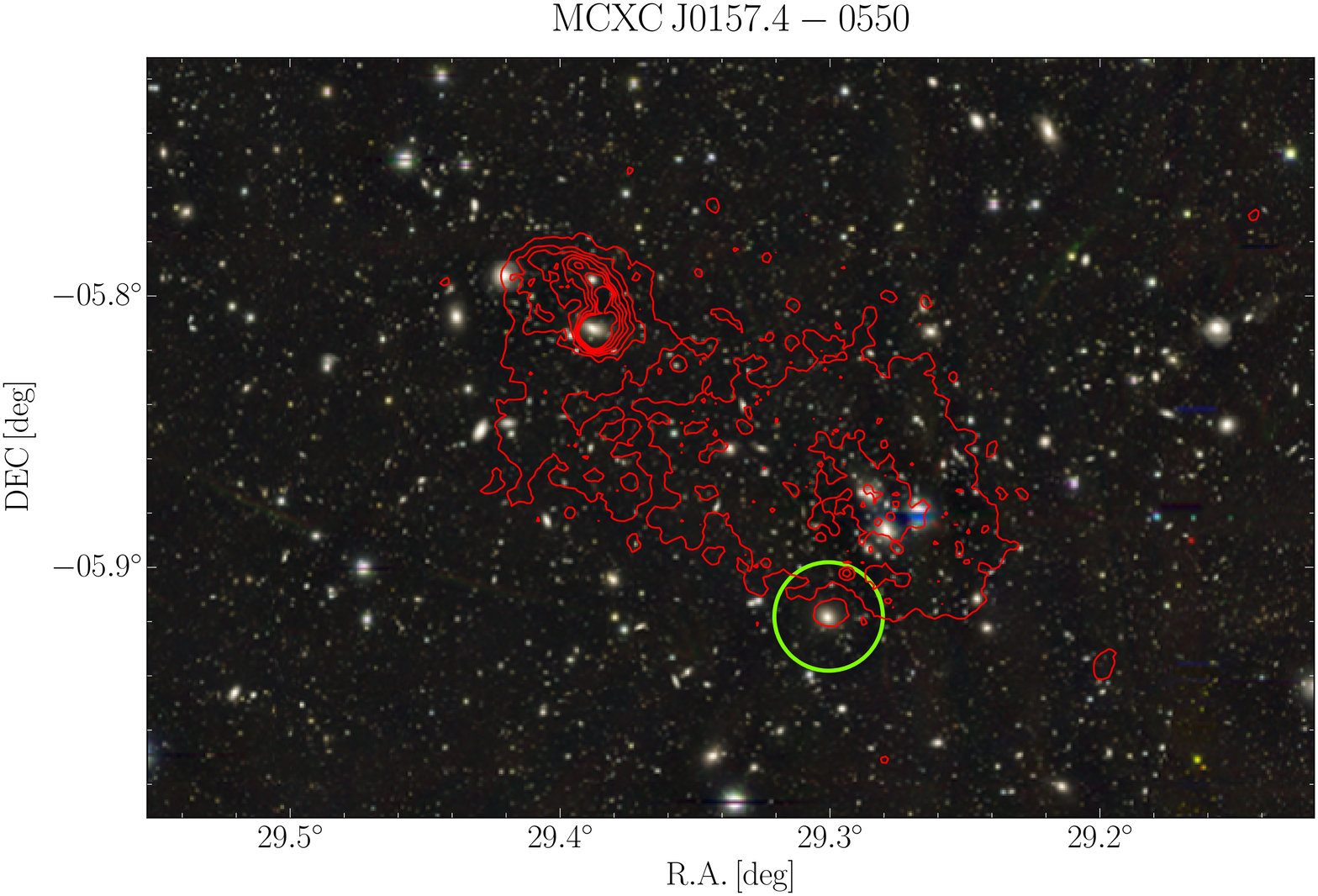}
\end{center}
\end{minipage}
\begin{minipage}{0.5\hsize}
\begin{center}
   \includegraphics[width=\hsize]{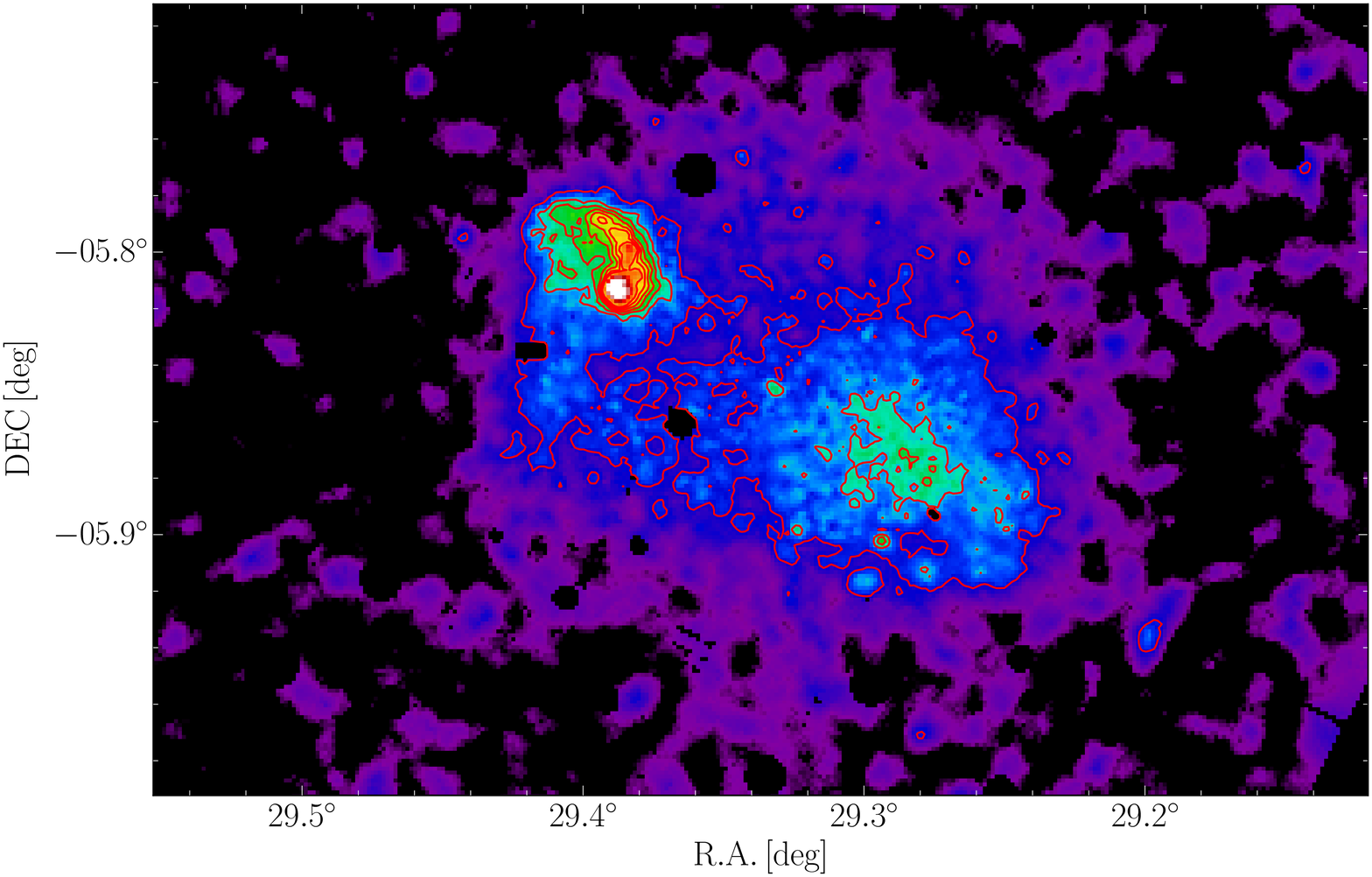}
\end{center}
\end{minipage}
\caption{Left : HSC $gri$ pseudo colour image for MCXCJ0157.4-0550, 
 overlaid with vignetting-corrected X-ray contours excluding point sources in red colour. The
 green circle has a $1.'2$ radius and is centered on the CAMIRA
 cluster. 
The BCG is located at $(\alpha,\delta)=(29.279,-5.887)$.
Right:
 Adaptively smoothed image in $0.4-2.3$keV excluding point
 sources.  The contours with the vignetting correction are 7 levels linearly spaced from [10-100] ${\rm cts~s}^{-1}{\rm deg}^{-2}$.
The northeast gas substructure clearly shows a comma-shaped feature,
 suggesting ram-pressure stripping.
}\label{fig:mcxcj0157}
\end{figure*}

MCXCJ0157.4-0550 is an on-going cluster merger at $z=0.1289$ as shown
in Figure \ref{fig:mcxcj0157}.
The system is composed of the western main gas halo and the
northeast subhalo. 
Optically-luminous galaxies are concentrated in the western region where
the X-ray morphology is highly elongated along the west-east direction.
The CAMIRA center is slightly offset from both X-ray and the BCG centers, 
because some luminous galaxies are missing in the HSC CAMIRA catalog. 
A comma-shaped feature in the X-ray emission is discovered at
$\sim1.6r_{500}\sim r_{200}$ ( $r_{200}$ based on H.E. mass
estimation (Section \ref{subsec:MHE})), which suggests ram pressure stripping of the sub-cluster.
An optically-luminous galaxy is located at the sub-cluster X-ray peak.
The redshift retrieved from the SDSS, $0.1286$, is very close
to the cluster redshift, suggesting that the substructure is infalling
in the plane of the sky. 
It is consistent with the fact that a prominent tail is observed.
The comma-shaped feature also implies that the infalling gas observed in
the X-ray has large angular-momentum.
If we improperly treat the substructure in the X-ray surface brightness ($S_x$) modeling, 
it may change the outer density slope (Equation \ref{eq:beta}) of the main halo
and eventually affect the H.E. mass estimation. 
Furthermore, we cannot rule out a possibility that the slopes in
the disturbed (east) and undisturbed (west) sectors are different.
%Since our analysis assumes spherical symmetry to measure spherical
%H.E. masses,
Since our analysis measures azimuthally-averaged X-ray quantities to
estimate spherical H.E. masses under the assumption of spherical symmetry,
it is not good to divide into azimuthally-dependent sub-sectors in
order to estimate the azimuthally-averaged outer slope.
To solve these problems, we implement the subtraction using the
off-centering effect (Equation \ref{eq:Sxoff}) in a model of the azimuthally averaged surface brightness profile
centering the western gas halo.
As a first approximation, we assume the
$\beta$ model for the western main gas halo and the northeast subhalo. The resulting profile well describes the observed surface
brightness (Figure \ref{fig:sx_mcxcj0157}). 
 
The H.E. mass is estimated only for the main cluster. 
The total gas mass within $\sim r_{200}$ computed for both the main gas and the gas substructure is $\sim8\%$ higher
than that estimated by the main cluster component, while the gas mass within
$r_{500}$ does not change.

Unfortunately, this cluster is located outside of the full-depth and
full-colour region of the HSC-SSP S16A data,
and thus the WL shape data are not available. 
A mass comparison for this cluster will be carried out in a future study.

\subsection{MCXCJ0231.7-0451}  \label{subsec:MCXCJ0231}

\begin{figure}
\begin{center}
   \includegraphics[width=\hsize]{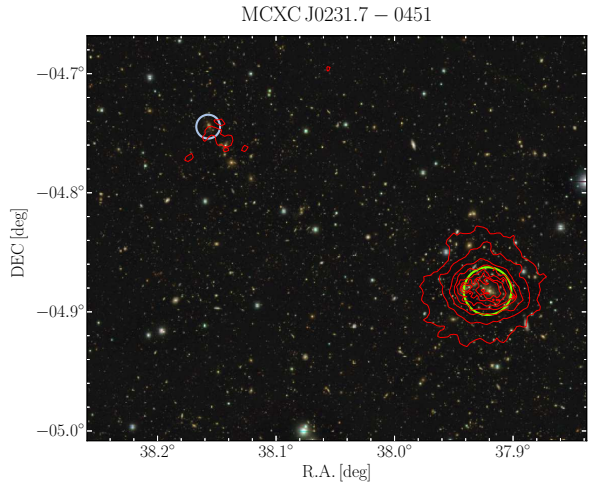}
\end{center}
\caption{Optical image of MCXCJ0231.7-0451, overlaid with
 vignetting-corrected X-ray contours in red
 colour. Green circles show $1.'2$ radius
 centering CAMIRA clusters. The western diffuse X-ray emission is from the target cluster.
 Faint X-ray emission from another CAMIRA cluster with $N_{\rm cor}=33.70$ 
 at ($29.203, -6.119$) and $z=0.2760$, marked by a light-blue circle
 with $0.'6$ radii, is also detected around the edges of the
 instruments. The contours with the vignetting correction are 7 levels linearly spaced from
 [10-100] ${\rm cts~s}^{-1}{\rm deg}^{-2}$.
}\label{fig:mcxcj0231}
\end{figure}

We %newly observed
present new observations of the cluster ($z=0.1843$) located in the XXL survey region with {\it
XMM-Newton}. An X-ray luminous point source at 
$\sim3.'5$ east of the X-ray center is found. Faint X-ray emission from another CAMIRA
cluster at $z=0.2760$ is also detected around the edge of the X-ray
detectors (Figure \ref{fig:mcxcj0231}). These X-ray sources are excluded in our analysis.

The cluster is referred to as XXL091 in the XXL survey \citep{Eckert16}, 
and has $M_{\rm gas}=5.00^{+0.80}_{-0.83}\times10^{13}\h70Msol$ within  
$r_{500}^{\rm MT}=1149\pm161h_{70}^{-1}{\rm kpc}$.
Here, $r_{500}^{\rm MT}$ is
derived from a scaling
relation between WL mass and X-ray temperature \citep{Lieu2016}.
We find that our measurement $M_{\rm
gas}=5.10^{+0.85}_{-0.84}\times10^{13}\h70Msol$ 
within the same radius is in good agreement with \cite{Eckert16}.

The mass estimation of the {\it Planck} SZ observation
\citep{Planck15CluterCosmology} gives  $M_{500}^{\rm
SZ}=3.96^{+0.49}_{-0.49}\times10^{14}\h70Msol$, 
which agrees with our H.E. mass estimate $M_{500}^{\rm
H.E.}=3.43^{+0.77}_{-0.65}\times10^{14}\h70Msol$.

Our WL mass measurement gives, $M_{500}^{\rm
WL}=7.96_{-1.89}^{+2.58}\times10^{14}\h70Msol$, which
agrees within errors with the CFHT WL mass measurement $M_{500}^{{\rm
WL}}=6.2^{+2.1}_{-1.8}\times10^{14}\h70Msol$ \citep{Lieu2016}.

\subsection{MCXCJ0201.7-0212}  \label{subsec:MCXCJ0201}

This cluster is known as Abell 291, and has a cool core \citep{Okabe10b,Okabe16,Martino14}.
Since the HSC-SSP S16A data of the cluster do not satisfy the
 full-depth and full-colour conditions for WL mass
measurement, we carry out X-ray and optical analysis.

\begin{figure}
\begin{center}
   \includegraphics[width=\hsize]{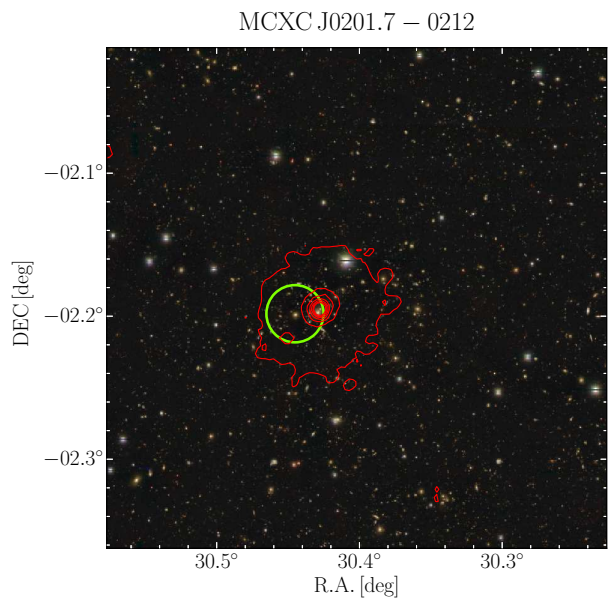}
\end{center}
\caption{MCXCJ0201.7-0212. The colours are the same as those in Figure
 \ref{fig:mcxcj0157}. The contours with the vignetting correction are 7 levels linearly spaced from [10-1000] ${\rm cts~s}^{-1}{\rm deg}^{-2}$.}\label{fig:mcxcj0201}
\end{figure}

We analyzed the same X-ray data used in \citet{Martino14}.
The gas density profile is well described by a double $\beta$ model. 
The CAMIRA center is slightly offset from the X-ray centroid and the BCG
(Figure \ref{fig:mcxcj0201}), because a few bright galaxies are missing
in the CAMIRA catalog.
Our H.E. mass estimate, $M_{500}^{\rm H.E.}=3.21_{-0.44}^{+0.51}\times10^{14}\h70Msol$, agrees with
a previous X-ray study $M_{500}=2.92\pm0.56\times10^{14}\h70Msol$
\citep{Martino14}.
These H.E. mass estimates are slightly lower than the corresponding estimated WL mass
$M_{500}^{\rm WL}=4.46^{+1.01}_{-0.96}\times10^{14}\h70Msol$ \citep{Okabe16b}, but the
difference is not statistically significant. 
A mass comparison using the HSC-SSP data is left for future works.

\subsection{MCXCJ1415.2-0030 and MCXCJ1415.2-0030W} \label{subsec:MCXCJ1415}

\begin{figure}
\begin{center}
   \includegraphics[width=\hsize]{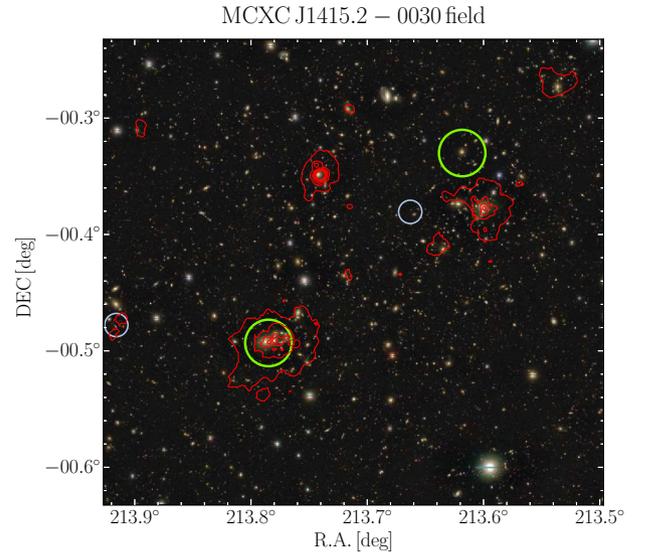}
\end{center}
\caption{MCXCJ1415.2-003 field, overlaid with vignetting-corrected X-ray contours in red
 colour. Four diffuse X-ray sources are found in this
 field. MCXCJ1415.2-0030 and MCXCJ1415.2-0030W are at middle-left and
 middle-right of the panel, respectively. The other two diffuse X-ray
 emissions surrounding the system are found at
 $(\alpha,\delta)=(213.740,-0.350)$ and
 $(\alpha,\delta)=(213.541,-0.272)$.  
The green circles have a $1.'2$ radius and are centered on the CAMIRA clusters. 
Two high-z CAMIRA clusters ($z>0.9$ and $N_{\rm cor}\simgt 15$) are
 found in the field, marked by light-blue circles with $0.'6$
 radii. The contours with the vignetting correction are 7 levels linearly spaced from [10-100] ${\rm cts~s}^{-1}{\rm deg}^{-2}$.}\label{fig:mcxcj1415}
\end{figure}

The system is mainly composed of the originally-identified MCXC cluster,
MCXCJ1415.2-0030, at the east and its companion cluster at $\sim2$ Mpc northwest of the MCXC cluster.  
We refer to the western cluster as MCXCJ1415.2-0030W for convenience. 
The X-ray emission shows no evidence for disturbance due to merger activity.
Besides the two clusters, two faint, diffuse X-ray emissions are found in the field (Figure \ref{fig:mcxcj1415}). 
The northern emission $(\alpha,\delta)=(213.740,-0.350)$ and the north-western
emission $(\alpha,\delta)=(213.541,-0.272)$ are associated with galaxies at $z=0.1389$ and $z=0.1398$,
respectively. These components of which radii are $\sim 0.'7-0.'8$ are excluded in the following X-ray analysis.
In the CAMIRA catalog, those galaxies are identified as a part of
MCXCJ1415.2-0030W, giving a large richness.
The X-ray emission from the eastern MCXC cluster
coincides with the CAMIRA center, 
while the western emission is $\sim3'$ offset from the CAMIRA center.
This is because the western CAMIRA cluster includes the northern and north-western groups.
We find no evidence that the BCGs of the eastern and western clusters
are significantly offset from the X-ray centroids.
The X-ray luminosity of the eastern MCXC cluster is
brighter than that of the western cluster, while the richness for the
western cluster is higher. 
\citet[][Figure 4 therein]{Owers13} have shown based on spectroscopic data that member
galaxies of the eastern and western clusters are spread over
$\sim4$~Mpc and $\sim1$~Mpc, respectively.

%We used two X-ray data pointed to the two clusters for measuring
We analyzed X-ray data for these two clusters in order to measure
 gas temperatures and surface brightness profiles. 
To carefully estimate density outer-slopes, 
we computed two surface brightness profiles centering on each of the two clusters 
and simultaneously fit them with the two surface brightness models with
the off-centering effect. 
We found that the observed profiles are well-described by the sum of X-ray
emission of the two clusters, requiring no extra component such as a 
filamentary gas component bridge between the two clusters.
In the surface brightness profile centered on MCXCJ1415.2-0030,
the flux of the cluster, the other cluster and the background at $R\sim1~{\rm Mpc}$ 
account for $\sim1\%$, $\sim13\%$ and $\sim86\%$, respectively. 
In the temperature measurements for each cluster,
we selected the background-dominated region for the annulus. Again, if
we ignore the flux contamination from the other cluster in the surface
brightness modelling, we overestimate the background component and
eventually misestimate the outer slopes.

Based on the H.E. assumption, we estimate 
$M_{500}^{\rm H.E.}=1.54_{-0.23}^{+0.34}\times10^{14}\h70Msol$ for MCXCJ1415.2-0030 and 
$M_{500}^{\rm H.E.}=0.44_{-0.07}^{+0.07}\times10^{14}\h70Msol$ for
MCXCJ1415.2-0030W (Table \ref{tab:mass}), respectively. 
It suggests that the originally-identified MCXC cluster is the main cluster.

We also carry out WL mass measurements for the two clusters.
We adopt a maximum radius for each tangential shear profile centered on 
each BCG of $\sim1.3h_{70}^{-1}$Mpc. Since the maximum radius is much less than the projected distance
between the two clusters, the off-centering effect of lensing signal is
negligible, $\sim {\mathcal O}(10^{-5})\times \langle \Sigma_{\rm cr}^{-1}\rangle^{-1}$ \citep{Yang06}, 
in contrast to X-ray analysis. 
This is caused by the fact that $\Delta \Sigma_{+,\rm off}=\bar{\Sigma}_{\rm
off}-{\Sigma}_{\rm off}$,
where the the surface mass density for the off-centering component at
the measured radius ($\Sigma_{\rm off}(R)$) is comparable to the mean surface mass density
within the radius ($\bar{\Sigma}_{\rm off}(<R)$). 
The WL masses are $M_{500}^{\rm
WL}=2.09_{-0.90}^{+1.43}\times10^{14}\h70Msol$, for MCXCJ1415.2-0030 and 
$M_{500}^{\rm WL}=0.80_{-0.58}^{+0.87}\times10^{14}\h70Msol$
MCXCJ1415.2-0030W (Table \ref{tab:mass}), respectively. 
Since the signal-to-noise ratio of the tangential shear profile for
MCXCJ1415.2-0030W is small, we used one parameter,
$M_{500}$, assuming the halo concentration based on the median value of
the mass versus concentration relation \citep{Diemer14}.
A sum of best-fit viral radii $\sim2.6~{\rm Mpc}$ is comparable to the
projected separation $\sim2~{\rm Mpc}$ and non-disturbed gas distribution,
suggesting that the two clusters are at early phase of cluster merger.
The H.E. mass for the western companion cluster is comparable to the WL mass.

Virial mass estimation \citep{Owers13} using spectroscopic data has shown that $M_{500}^{\rm
vir}=1.5\pm0.3\times10^{14}\h70Msol$ for the main cluster (A1882A) and $M_{500}^{\rm
vir}=1.0\pm0.5\times10^{14}\h70Msol$ for the companion cluster (A1882B), 
respectively. Dynamical mass estimates are in good agreement with our WL masses. 
\cite{Owers13} also concluded based on joint X-ray and kinematics
analysis that the system is likely to be before cluster merger.
Our results agree with their conclusions.

\subsection{Richness vs $M_{\rm H.E.}$}

We compare the H.E. masses for the MCXC clusters with the CAMIRA cluster richness. 
Since the cluster richness is generally proportional to the number of
member galaxies, it is expected to be a good mass
proxy. Indeed, \cite{Oguri17} have compared public X-ray
temperatures and luminosities in the XXL and XMM-LSS fields with their CAMIRA
cluster richnesses, and found good correlations.
The slope in the richness and temperature scaling relation
\citep{Oguri17} is
found to be shallower than that predicted by a self-similar solution.
The temperatures are measured within a fixed radius, $300$ kpc, 
and thus are potentially and partially affected by baryonic physics. 
We here study a correlation between the H.E. masses and cluster richness.

\begin{figure}
\begin{center}
   \includegraphics[width=\hsize]{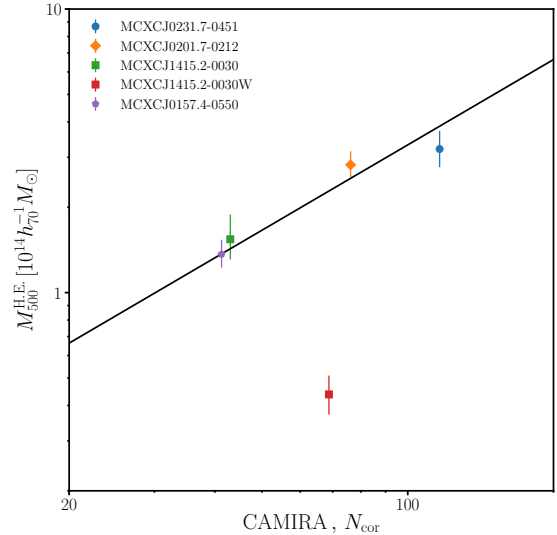}
\end{center}
\caption{A comparison between CAMIRA richness \citep{Oguri17} and
 HE mass at $\Delta=500$. A solid line is the best-fit for the four MCXC
 clusters. The H.E. mass for the non-MCXC cluster is significantly lower than the best-fit base
 line.}\label{fig:mhe-Ncor}
\end{figure}

Figure \ref{fig:mhe-Ncor} compares the richness with the H.E. mass. 
The H.E. masses for the original sample of the MCXC clusters are
$M_{500}\gtsim 1.5\times10^{14}\h70Msol$, which is consistent with
the masses we expected from choosing high luminosity clusters for
this study, 
$M_{500}\gtsim 2\times10^{14}\h70Msol$ (Section \ref{sec:HSCSSP}).
A small discrepancy is acceptable when we consider intrinsic scatter
in the scaling relation \citep{Piffaretti11}.
We fit the relation with a signal-power law model
\begin{eqnarray}
\log \left(\frac{M_{500}^{\rm H.E.}}{10^{14}\h70Msol}\right)=a \log
 N_{\rm cor} +b 
\end{eqnarray}
We here consider the relation consistent with our measurements for the four MCXC clusters
and obtain 
$a=0.84\pm0.15$ and $b=-2.73\pm0.61$.
The best-fit slope agrees with  $\sim 1$ predicted by $N_{\rm
cor}\propto M$ \citep{Lin04}. 
The best-fit normalization suggests that $M_{500}^{\rm
H.E.}\sim6\times10^{13}\h70Msol$ at $N_{\rm cor}=15$. 
When we fix $a=1$, we obtain 
$b=-3.41\pm0.50$ and $M_{500}^{\rm H.E.}\sim5\times10^{13}\left(N_{\rm cor}/15\right)\h70Msol$.
\cite{Oguri17} have shown that $N_{\rm cor}=15$ roughly corresponds to
$M_{200\rm m}\sim10^{14}\hMsol$, if the number of discovered clusters
agrees with the prediction of a cluster mass function computed by
\cite{Tinker10} with $\sigma_8=0.82$. 
Here, $200\rm m$ means that the mean density is 200 times the mean matter density of the
Universe.
Assuming the median halo concentration $c_{200\rm m}=6$ \citep{Diemer14}, $M_{200\rm
m}\sim10^{14}\hMsol$ gives $M_{500}\sim7\times 10^{13}\h70Msol$. 
Our H.E. mass estimation roughly agrees with the expectations of
\cite{Oguri17}.
More precise comparison using a large number of clusters will be carried
out in future works.

Interestingly, the H.E. mass for the non-MCXC cluster,
MCXCJ1415.2-0030W, is significantly lower than the best-fit base line.
The deviation might be explained by two possibilities or their
combination. First, at the early stage of cluster merger, the ICM would strongly deviate
from H.E., consistent with our finding that the WL mass is higher (Section \ref{subsec:MCXCJ1415}). 
Second, the richness would be overestimated because the CAMIRA member
galaxies of MCXCJ1415.2-0030W include the other group components.
We also fit the mass-richness scaling relation for the five clusters
with $a=1$ fixed, and confirm that the normalization, $b=-3.74\pm0.45$ %$b=-3.70\pm0.45$
does not significantly change.

\subsection{Mass comparison} \label{sec:mass_comp}

\begin{figure}
\begin{center}
   \includegraphics[width=\hsize]{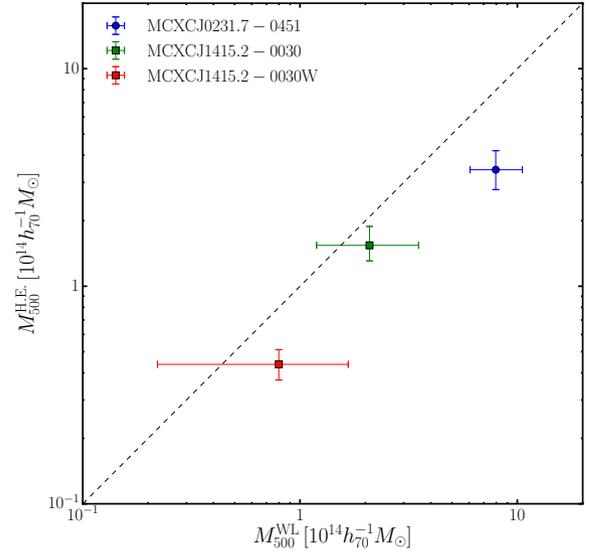}
\end{center}
\caption{Mass comparison of WL and H.E. masses for the three clusters
 at $\Delta=500$.}\label{fig:mhe-mwl2}
\end{figure}

We compare WL masses with H.E. masses at $\Delta=500$, as a first
attempt of our further studies.
The full-depth and full-colour
conditions allows us to compare the two masses only for the three
clusters. The X-ray and WL mass measurements are described in Sections
\ref{subsec:MHE} and \ref{sec:MWL}, respectively. 
As in previous studies of hydrostatic mass bias
\citep{Mahdavi13,vonderLinden14,Okabe14b,Hoekstra15,Smith16}, 
a comparison of H.E. and WL mass gives an indirect constraint on 
the degree to which the assumption that clusters are in H.E. is valid
for cosmological applications. 
We deliberately perform a simpler calculation of mass bias because of the observational limitation
of the current sample. 
The mass comparison adopts the masses enclosed within the overdensity radii independently
determined by different measurements. When we compare the masses
measured within the same apertures, the results do not change and thus
the aperture mismatching is a subdominant effect. 
We adopt the unweighted geometric mean to quantify the mass bias $b_m$, 
\begin{eqnarray}
% b_m=1-\exp \left[\frac{1}{n}\sum_{i}\ln\left(\frac{M_{\rm H.E.}}{M_{\rm
%					 WL}}\right)_i\right]. 
 b_m=1-\prod_{i}^{n} \left(\frac{M_{\rm H.E.}}{M_{\rm WL}}\right)_i^{1/n}.
\end{eqnarray}
When we exchange $M_{\rm H.E.}$ for $M_{\rm WL}$ in the equation, the
second term of this quantity becomes the inverse 
in contrast to an estimation with $\sum_i \left(M_{\rm H.E.}/M_{\rm WL}\right)_i/n$.
If the H.E. mass is statistically consistent with the WL mass, the bias
parameter, $b_m$ is equal to zero.  
We obtain %$b_m=0.43^{+0.31}_{-0.45}$,
$b_m=0.44_{-0.45}^{+0.31}$,
for the three clusters at $\Delta=500$ (Figure \ref{fig:mhe-mwl2}). 
We also obtain %$b_m=0.33_{-0.17}^{+0.14}$
$b_m=0.34_{-0.19}^{+0.16}$ for the two MCXC clusters. 
Since the measurement errors for WL masses of the two MCXC clusters are 
relatively small, the error of $b_m$ becomes smaller.
It indicates that the H.E. mass at $\Delta=500$ is consistent with
the WL mass in the current sample at the $2\sigma$ level.

When we use the same radii determined by weak-lensing masses,
we obtain $b_m=0.40_{-0.49}^{+0.39}$ for the three clusters and
$b_m=0.34_{-0.41}^{+0.28}$ for the two clusters, respectively. We here
consider the error propagation of measurement uncertainties of the WL
radii. The result does not significantly change.
When we measure WL masses with X-ray centers, the best-fit WL masses are changed
only by a few percent because the offset distances between X-ray
centroids and BCG positions are very small.

Although our results are statistically poor because of a small
sample of clusters, we compare with the
literature. Direct comparisons between weak-lensing masses and X-ray masses are
not trivial, because previous studies applied their own methods: 
the boost factor correction \citep[e.g.][]{vonderLinden14,Hoekstra15} or no
correction \citep[e.g.][]{Okabe16,Umetsu16} in WL analyses and emission-weighted temperatures  \cite[e.g.][]{Zhang08, Mahdavi13} or
spectroscopic-like temperature \citep[e.g.][]{Mazzotta04,Martino14} in
X-ray analyses. 
\cite{Smith16} obtained the average bias $b_m=0.05\pm0.05$ for fifty clusters at
$z\sim0.2$, and \cite{Mahdavi13} computed $b_m=0.12\pm0.05$ with their WL radii.
Using the same sample between the two papers, the major difference
($\sim10\%$) would come from X-ray mass measurements \citep{Smith16}.
We here assume that the difference is mainly caused by temperature definitions
and discuss this possibility.  
\cite{Mazzotta04} discovered using realistic simulations that the
H.E. mass estimations with emission-weighted temperatures would be
underestimated by $\sim10\%$ and those with spectroscopic-like
temperature would recover the input mass. 
When we estimate the H.E masses with emission-weighted temperatures, the
masses are indeed lower $\sim10\%$ than our results.
Therefore, the possibility does not conflict with a difference between
the two papers \citep{Mahdavi13,Smith16}.
However, the current uncertainty for the averaged bias is too large to
discuss the details. 
When we compile the full sample of $22$ clusters,
we expect that the uncertainty for the average bias will be comparable to those for previous
studies for $50$ clusters \citep{Hoekstra15,Smith16}.
We therefore will compare WL and H.E. masses for the full sample, and
investigate the redshift dependence and radial dependence of the mass bias.

\subsection{Baryon Fraction} \label{subsec:fb}

The ratio of baryonic-to-total mass in massive clusters 
is expected to closely match the cosmic mean baryon
fraction, $\Omega_b/\Omega_m$ measured from CMB experiments if baryons are trapped in potential wells \citep[e.g.][]{Evrard97,Kravtsov05}.
However, the baryon budget in galaxy clusters is sensitive to
non-gravitational process; stars are formed from gas
%stellar masses are converted from gas masses
through radiative cooling and AGN feedback may push the gas outside the potential well. 
Thus, measurements of the cluster baryon fraction are important to 
understand baryonic physics and the interplay between baryons and dark matter. 
Furthermore, assuming that the gas mass fraction is constant across
redshifts, gas mass fraction measurements potentially provide a
cosmological probe \citep[e.g.][]{Allen08}.

\begin{figure}
\begin{center}
   \includegraphics[width=\hsize]{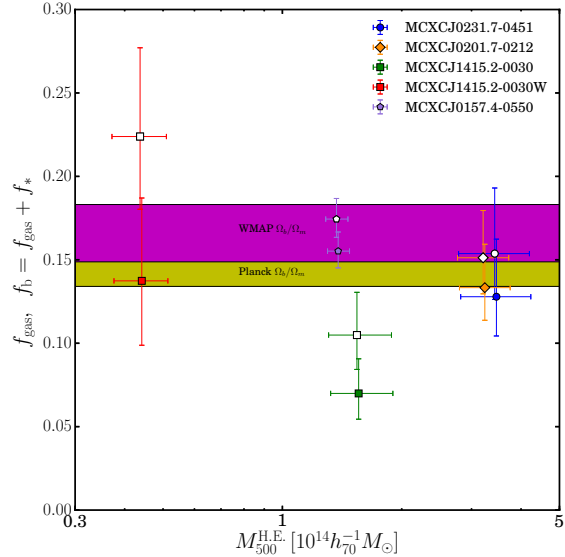}
\end{center}
\caption{Gas fraction (filled symbols), $f_{\rm gas}$, and baryon
 fraction (opened symbols), $f_{\rm b}$, within $r_{500}$ based on H.E. masses. 
 The horizontal filled regions are the cosmic mean baryon
 fraction $\Omega_b/\Omega_m$ for {\it WMAP} \citep{WMAP09} and {\it
 Planck} \citep{Planck15Cosmology} with their respective $1\sigma$
 uncertainties. For a visual purpose, the H.E. masses of the gas
 fractions are multiplied by $1.01$.
}\label{fig:fgas}
\end{figure}

This paper focuses on the baryon fraction at $\Delta=500$ based on the
H.E. mass, because the result based on the WL mass is statistically poor.
We define gas and baryon fraction as follows:
\begin{eqnarray}
%f_{\rm gas}(<r)&=&M_{\rm gas}(<r)/M_{\rm H.E.}(<r) \\ \nonumber
%f_{\rm b}(<r)&=&(M_{\rm gas}(<r)+M_{\rm *}(<r))/M_{\rm H.E.}(<r). \nonumber
f_{\rm gas}(<r)&=& \frac{M_{\rm gas}(<r)}{M_{\rm H.E.}(<r)}, \nonumber \\ 
f_{\rm b}(<r)&=& \frac{M_{\rm gas}(<r)+M_{\rm *}(<r)}{M_{\rm
 H.E.}(<r)}. \nonumber
\end{eqnarray}
 Here, $M_{\rm gas}$, $M_*$ and $M_{\rm H.E.}$ are gas, stellar and
H.E. masses (Table \ref{tab:mass}), respectively.  Gas mass is measured from X-ray analysis.  
Stellar masses are delivered from the deprojection estimation of the
CAMIRA cluster catalog using the HSC-SSP five-band photometry
(Section \ref{sec:camira}). 
Measurement uncertainty of the total mass propagates through
the over-density radius into gas and stellar masses. 
Figure \ref{fig:fgas} shows the gas and baryon fractions based on H.E. masses.  

We also investigated how much the stellar mass estimation is
changed if blue galaxies are included. We selected blue galaxies of which colors are
bluer by $1-3\sigma$ than those of the red-sequence galaxies within
$r_{500}$ and estimated their stellar mass in a cylinder volume
subtracted by $(2-3)r_{500}$ as the background region.
The total stellar masses are changed only by sub-$\%$.
Even if we neither subtract the background components nor change the
background region, the result does not significantly change.
It is not
surprising because the faint and blue galaxies are not dominant contributors to the light or stellar mass in
cluster central regions, in contrast to the bright and red galaxies.
We note that the stellar mass estimation for blue galaxies is within the projected 
cylinder volume because the characteristic spatial distribution for the
blue galaxies (essentially a hollowed-out sphere) makes it impractical to carry out the deprojection method.
We stress that we estimated the total stellar mass using red galaxies in
a spherical volume using the deprojection method (Section \ref{sec:camira}).

In contrast to previous observational studies
\citep[e.g.][]{Lin03,Vikhlinin09b} showing that gas mass fraction
increases and stellar mass fraction decreases with a total mass
increasing, 
we find no significant evidence of a halo mass dependence of
$f_{\rm gas}$ and $f_*$ in the current sample. The relation might be
difficult to measure given the intrinsic scatter and the small sample size.
We therefore focus on a comparison between the averages for $f_{\rm
gas}$ and $f_*$ for the current sample and the literature. 
Based on the defined selection function of the MCXC clusters, we compute
unweighted averages of gas and baryon fractions for the four MCXC clusters.

\begin{figure*}
\begin{minipage}{0.5\hsize}
\begin{center}
   \includegraphics[width=\hsize]{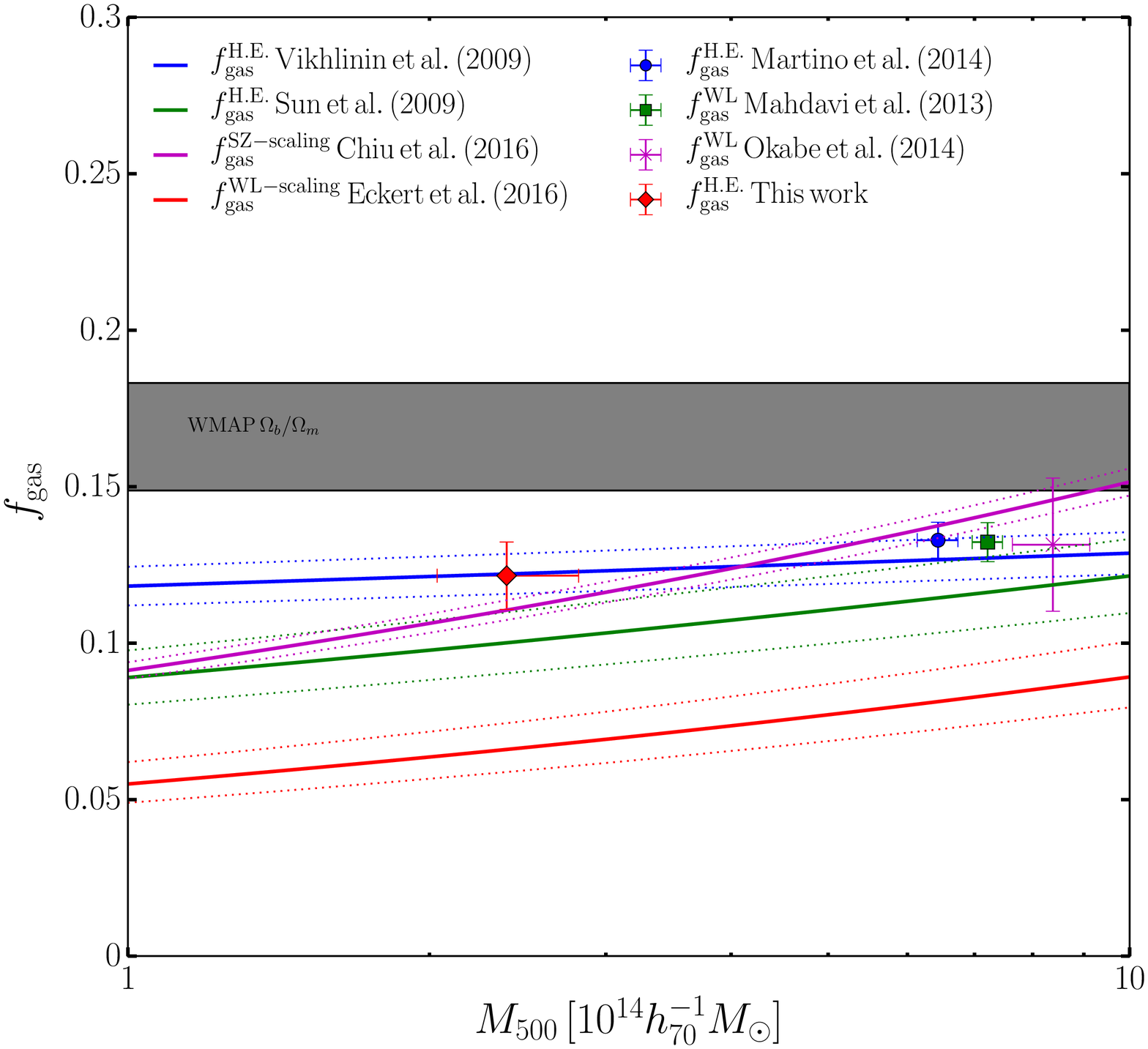}
\end{center}
\end{minipage}
\begin{minipage}{0.5\hsize}
\begin{center}
   \includegraphics[width=\hsize]{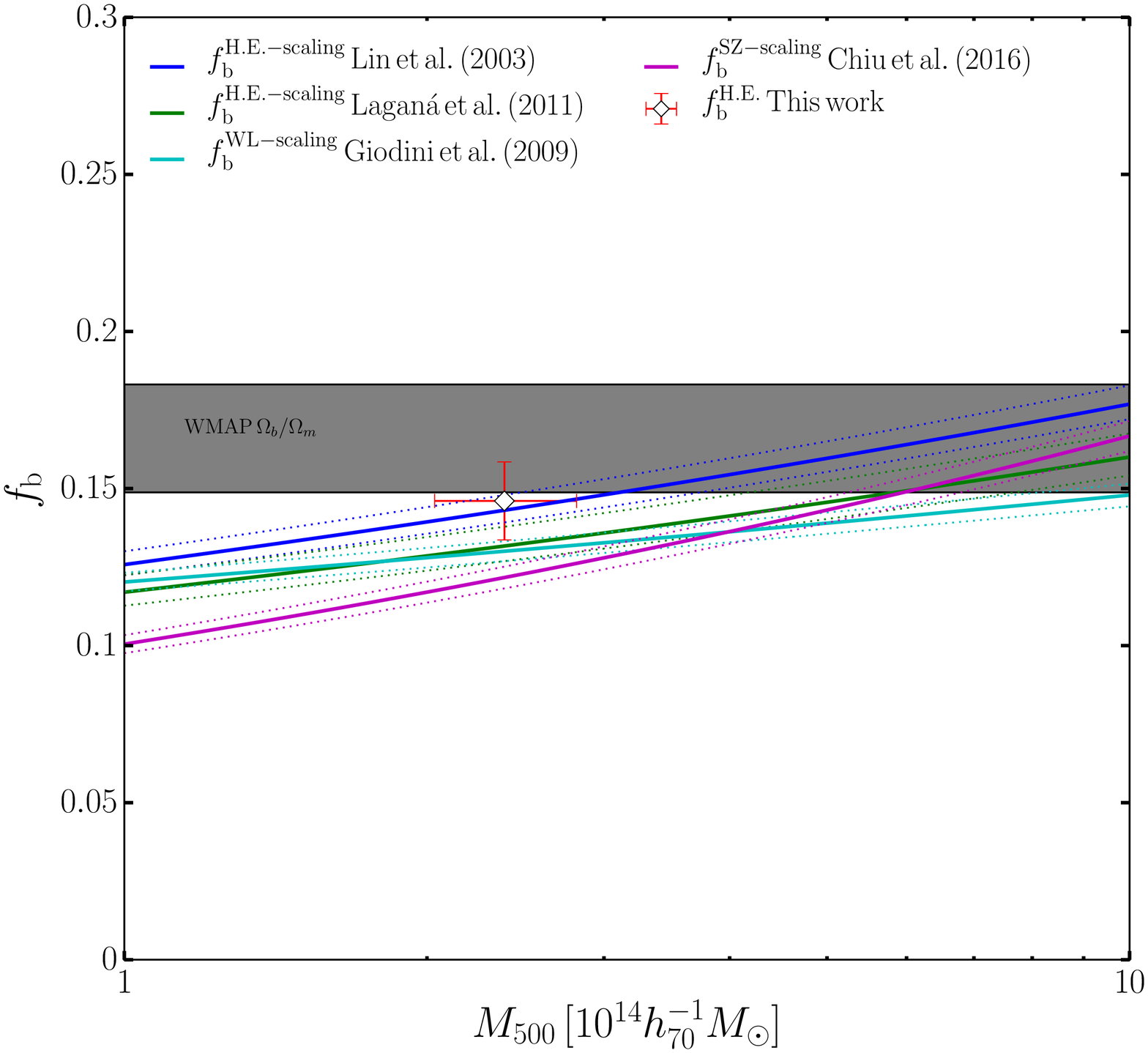}
\end{center}
\end{minipage}
\caption{Comparison with the literature : gas fraction (left) and baryon
 fraction (right). Left : Red diamond is the
 average for $f_{\rm gas}$ for the four MCXC clusters.
Blue circle, green square and magenta cross are $\langle f_{\rm gas} \rangle$ from \cite{Martino14},
 \cite{Mahdavi13} and \cite{Okabe14b}, respectively. 
Blue, green, magenta and red solid lines are scaling functions between gas fraction and mass from
\cite{Vikhlinin09a}, \cite{Sun09}, \cite{Chiu16}, and \cite{Eckert16},
 respectively. Dotted lines are $1\sigma$ uncertainties of the scaling functions. 
For simplicity, we plot $\Omega_b/\Omega_m$ for {\it WMAP} \citep{WMAP09}. Upper superscripts denote methods of
 total mass estimation.
Right : White diamond are the average for $f_{\rm b}$ for the four MCXC clusters,
 derived by this study. 
Blue, green, cyan and magenta solid lines are scaling functions between baryon fraction and mass from
 \cite{Lin03}, \cite{Lagana11}, \cite{Giodini09}, and
 \cite{Chiu16}. respectively. 
}\label{fig:fb_others}
\end{figure*}

To investigate a mass dependence of $f_{\rm gas}$ using the literature, 
we plot the averaged fraction enclosed within $r_{500}$ and mass plane
(left panel of Figure \ref{fig:fb_others}). 
The average value is $\langle f_{\rm gas} \rangle =0.125\pm0.012$,
%$\langle f_{\rm gas} \rangle =0.126\pm0.010$,
which is in agreement with previous
studies based on H.E. mass or Sunyaev-Zel'dovich effect (SZE) mass
\citep[e.g.][]{Vikhlinin09a,Martino14,Sun09,Chiu16} and based on WL masses
\citep[e.g.][]{Zhang10,Mahdavi13,Okabe14b}.
All points are unweighted averages from tables in the literature. 
Differences for those gas fractions at $M_{500}\sim 
2.4\times10^{14}\h70Msol$ and $\sim7\times10^{14}\h70Msol$ are $\sim 8\%$
and $\sim 6\%$, respectively.
However, the gas fraction of the XXL survey \citep{Eckert16} is
systematically lower than in other studies in a wide mass range. 
The deviation is at the $\sim5.3\sigma$ level, where we use the $8\%$ scatter.
In our sample of the four MCXC clusters, the gas mass fraction is
$\sim0.8\pm0.1$ of the cosmic mean baryon fraction $\Omega_b/\Omega_m$
for {\it WMAP} \citep{WMAP09} and $\sim0.9\pm0.1$ for {\it  Planck}
\citep{Planck15Cosmology}, though the two experiments have reported slightly discrepant results.
The values are slightly higher than $f_{\rm
gas}\Omega_m/\Omega_b\sim0.6$ from numerical simulations \citep[e.g.][]{Kravtsov05,Planelles13,Battaglia13}.

The average baryon fraction for the four MCXC clusters, $\langle f_{\rm b} \rangle =0.146\pm0.012$,
%$\langle f_{\rm b} \rangle =0.150\pm0.011$,
is comparable to $\Omega_b/\Omega_m$ (right panel of Figure \ref{fig:fb_others}).
Our result is also comparable to previous observational studies
\citep{Lin03,Lin12, Giodini09, Lagana11,Chiu16}.
A difference between those baryon fractions at $M_{500}\sim
2.4\times10^{14}\h70Msol$ is only $\sim 7\%$.
There are some discrepancies in $f_b$ even between different numerical simulations.
\cite{Kravtsov05} have shown that the total baryon fraction agrees with
$\Omega_b/\Omega_m$, while \cite{Planelles13} have pointed out that it
accounts for $\sim85\%$ because some fraction of gas is displaced
outside potential wells by AGN activities.

We also note that, if there were H.E. mass bias, the gas and baryon fractions
would be overestimated. 
Observations of baryon budget in galaxy clusters are still open questions.
% at least less massive clusters and groups, are still open questions.
Since small clusters and groups are sensitive to baryonic physics 
\citep[e.g.][]{Kravtsov05,Planelles13,Battaglia13}, 
future progress of the HSC-SSP survey and future studies based on WL
masses will play a key role in this subject.

\section{Summary}

We selected X-ray luminous clusters from the MCXC cluster catalog
\citep{Piffaretti11} to measure H.E. masses for galaxy clusters in the HSC-SSP survey region.
Based on the {\it XMM-Newton} and HSC-SSP datasets, we carried out a multiwavelength study of four MCXC clusters in the
S16A field and a non-MCXC cluster associated with one MCXC cluster.

We found a correlation between cluster richness and
H.E. mass for the MCXC clusters. The mass normalization agrees with
expectations by comparing the CAMIRA cluster abundance with a
theoretical prediction of cluster mass function with $\sigma_8=0.82$ \citep{Oguri17}.  
However, an infalling cluster to one MCXC cluster is highly deviant from
the scaling relation, which could be caused by mass underestimation
and/or richness overestimation.
The average cluster gas mass fraction based on H.E. masses,
$\langle f_{\rm b} \rangle = 0.146\pm0.012$,
%$\langle f_{\rm gas} \rangle =0.126\pm0.010$,
accounts for $\sim80-90\%$ of the cosmic mean baryon fraction. 
%One major uncertainty comes from a discrepancy between {\it WMAP} and {\it Planck}. 
In comparison with gas and baryon fractions from the literature based on
various mass measurements
\citep{Vikhlinin09a,Martino14,Mahdavi13,Okabe14b,
Sun09,Eckert16,Giodini09,Chiu16}, 
our measurements are somewhat higher than previous studies but overall are agreed.
Differences of gas and baryon fractions between these studies are
$\sim8\%$ and $\sim7\%$ at $M_{500}\sim 2.4\times10^{14}\h70Msol$, respectively. 
We also note a possibility that the
average gas and baryon fraction is somehow overestimated if there were H.E. mass bias.
Therefore, future studies using WL masses for a large number of
clusters/groups will be important to understand the baryon budgets and
improve the current level of the quality.

The full-depth and full-colour conditions of the HSC-SSP survey allows us to
compare H.E. mass with WL mass for the three clusters. 
The estimated mass bias, $b_m=0.44^{+0.31}_{-0.45}$, allows for the possibility that the H.E. masses agree
with the WL ones. In order to quantify the validly of H.E. assumption, 
we need to carry out WL analysis for the full sample of clusters.

Further joint studies using a large number of clusters are vitally
important to improve statistical uncertainty. 
Pointed X-ray observations with {\it XMM-Newton} and {\it Chandra} with
sufficient integration times are
essential to fairly compare X-ray observables with WL and optical
measurements. The approach is complementary to the forthcoming X-ray
survey from {\it eROSITA}, whose typical exposure in the HSC-SSP survey region is
too shallow to estimate H.E. masses. A collaboration with the on-going XXL survey is powerful to
understand cluster physics and carry out cluster-based cosmology. 
In similar ways, joint studies with the ACTPol SZE observations \citep{Miyatake17} provide us with an unique route for cluster studies.
Future studies based on survey-type datasets will also reveal how much cluster properties are changed by
cluster selection methods, like X-ray, SZE, optical and WL techniques.
The paper has demonstrated the power and impact of the HSC-SSP survey on
other wavelengths and shown 
the first result as a series of multiwavelength studies.

\begin{ack}

The Hyper Suprime-Cam (HSC) collaboration includes the astronomical communities of Japan and Taiwan, and Princeton University. The HSC instrumentation and software were developed by the National Astronomical Observatory of Japan (NAOJ), the Kavli Institute for the Physics and Mathematics of the Universe (Kavli IPMU), the University of Tokyo, the High Energy Accelerator Research Organization (KEK), the Academia Sinica Institute for Astronomy and Astrophysics in Taiwan (ASIAA), and Princeton University. Funding was contributed by the FIRST program from Japanese Cabinet Office, the Ministry of Education, Culture, Sports, Science and Technology (MEXT), the Japan Society for the Promotion of Science (JSPS), Japan Science and Technology Agency (JST), the Toray Science Foundation, NAOJ, Kavli IPMU, KEK, ASIAA, and Princeton University. 

This paper makes use of software developed for the Large Synoptic Survey Telescope. We thank the LSST Project for making their code available as free software at  http://dm.lsst.org

The Pan-STARRS1 Surveys (PS1) have been made possible through contributions of the Institute for Astronomy, the University of Hawaii, the Pan-STARRS Project Office, the Max-Planck Society and its participating institutes, the Max Planck Institute for Astronomy, Heidelberg and the Max Planck Institute for Extraterrestrial Physics, Garching, The Johns Hopkins University, Durham University, the University of Edinburgh, Queen’s University Belfast, the Harvard-Smithsonian Center for Astrophysics, the Las Cumbres Observatory Global Telescope Network Incorporated, the National Central University of Taiwan, the Space Telescope Science Institute, the National Aeronautics and Space Administration under Grant No. NNX08AR22G issued through the Planetary Science Division of the NASA Science Mission Directorate, the National Science Foundation under Grant No. AST-1238877, the University of Maryland, and Eotvos Lorand University (ELTE) and the Los Alamos National Laboratory.

Based on data collected at the Subaru Telescope and retrieved from the HSC data archive system, which is operated by Subaru Telescope and Astronomy Data Center at National Astronomical Observatory of Japan.

This work was supported by the Funds for the Development of Human
Resources in Science and Technology under MEXT, Japan and 
 Core Research for Energetic Universe in Hiroshima University (the MEXT
 program for promoting the enhancement of research universities, Japan).
This work was supported in part by World Premier International
Research Center Initiative (WPI Initiative), MEXT, Japan.
This work was supported by MEXT KAKENHI No. 26800097(NO), 26800093/15H05892(MO), 
15K05080 (YF), 26400218 (MT) and 15K17610 (SU).
HM is supported by the Jet Propulsion Laboratory, California Institute
 of Technology, under a contract with the National Aeronautics and Space
 Administration.

The paper is dedicated to the memory of our friend, Dr. Yuying Zhang,
 who sadly passed away in 2016. She gave helpful suggestions on our
 X-ray analysis.

\end{ack}

\bibliographystyle{apj}
\bibliography{my,hsc,mcxc-hsc,mcxc-hsc_sub}

\begin{thebibliography}{105}
\expandafter\ifx\csname natexlab\endcsname\relax\def\natexlab#1{#1}\fi

\bibitem[{{Aihara} {et~al.}(2017{\natexlab{a}}){Aihara}, {Armstrong},
  {Bickerton}, {Bosch}, {Coupon}, {Furusawa}, {Hayashi}, {Ikeda}, {Kamata},
  {Karoji}, {Kawanomoto}, {Koike}, {Komiyama}, {Lupton}, {Mineo}, {Miyatake},
  {Miyazaki}, {Morokuma}, {Obuchi}, {Oishi}, {Okura}, {Price}, {Takata},
  {Tanaka}, {Tanaka}, {Tanaka}, {Uchida}, {Uraguchi}, {Utsumi}, {Wang},
  {Yamada}, {Yamanoi}, {Yasuda}, {Arimoto}, {Chiba}, {Finet}, {Fujimori},
  {Fujimoto}, {Furusawa}, {Goto}, {Goulding}, {Gunn}, {Harikane}, {Hattori},
  {Hayashi}, {Helminiak}, {Higuchi}, {Hikage}, {Ho}, {Hsieh}, {Huang}, {Huang},
  {Imanishi}, {Iwata}, {Jaelani}, {Jian}, {Kashikawa}, {Katayama}, {Kojima},
  {Konno}, {Koshida}, {Leauthaud}, {Lee}, {Lin}, {Lin}, {Mandelbaum},
  {Matsuoka}, {Medezinski}, {Miyama}, {Momose}, {More}, {More}, {Mukae},
  {Murata}, {Murayama}, {Nagao}, {Nakata}, {Niikura}, {Nishizawa}, {Oguri},
  {Okabe}, {Ono}, {Onodera}, {Onoue}, {Ouchi}, {Pyo}, {Shibuya}, {Shimasaku},
  {Simet}, {Speagle}, {Spergel}, {Strauss}, {Sugahara}, {Sugiyama}, {Suto},
  {Suzuki}, {Tait}, {Takada}, {Terai}, {Toba}, {Turner}, {Uchiyama}, {Umetsu},
  {Urata}, {Usuda}, {Yeh}, \& {Yuma}}]{HSC1styr}
{Aihara}, H., {Armstrong}, R., {Bickerton}, S., {et~al.} 2017{\natexlab{a}},
  ArXiv e-prints

\bibitem[{{Aihara} {et~al.}(2017{\natexlab{b}}){Aihara}, {Arimoto},
  {Armstrong}, {Arnouts}, {Bahcall}, {Bickerton}, {Bosch}, {Bundy}, {Capak},
  {Chan}, {Chiba}, {Coupon}, {Egami}, {Enoki}, {Finet}, {Fujimori}, {Fujimoto},
  {Furusawa}, {Furusawa}, {Goto}, {Goulding}, {Greco}, {Greene}, {Gunn},
  {Hamana}, {Harikane}, {Hashimoto}, {Hattori}, {Hayashi}, {Hayashi},
  {He{\l}miniak}, {Higuchi}, {Hikage}, {Ho}, {Hsieh}, {Huang}, {Huang},
  {Ikeda}, {Imanishi}, {Inoue}, {Iwasawa}, {Iwata}, {Jaelani}, {Jian},
  {Kamata}, {Karoji}, {Kashikawa}, {Katayama}, {Kawanomoto}, {Kayo}, {Koda},
  {Koike}, {Kojima}, {Komiyama}, {Konno}, {Koshida}, {Koyama}, {Kusakabe},
  {Leauthaud}, {Lee}, {Lin}, {Lin}, {Lupton}, {Mandelbaum}, {Matsuoka},
  {Medezinski}, {Mineo}, {Miyama}, {Miyatake}, {Miyazaki}, {Momose}, {More},
  {More}, {Moritani}, {Moriya}, {Morokuma}, {Mukae}, {Murata}, {Murayama},
  {Nagao}, {Nakata}, {Niida}, {Niikura}, {Nishizawa}, {Obuchi}, {Oguri},
  {Oishi}, {Okabe}, {Okura}, {Ono}, {Onodera}, {Onoue}, {Osato}, {Ouchi},
  {Price}, {Pyo}, {Sako}, {Okamoto}, {Sawicki}, {Shibuya}, {Shimasaku},
  {Shimono}, {Shirasaki}, {Silverman}, {Simet}, {Speagle}, {Spergel},
  {Strauss}, {Sugahara}, {Sugiyama}, {Suto}, {Suyu}, {Suzuki}, {Tait},
  {Takata}, {Takada}, {Tamura}, {Tanaka}, {Tanaka}, {Tanaka}, {Tanaka},
  {Terai}, {Terashima}, {Toba}, {Toshikawa}, {Turner}, {Uchida}, {Uchiyama},
  {Umetsu}, {Uraguchi}, {Urata}, {Usuda}, {Utsumi}, {Wang}, {Wang}, {Wong},
  {Yabe}, {Yamada}, {Yamanoi}, {Yasuda}, {Yeh}, {Yonehara}, \&
  {Yuma}}]{HSC1styrOverview}
{Aihara}, H., {Arimoto}, N., {Armstrong}, R., {et~al.} 2017{\natexlab{b}},
  ArXiv e-prints

\bibitem[{{Allen} {et~al.}(2008){Allen}, {Rapetti}, {Schmidt}, {Ebeling},
  {Morris}, \& {Fabian}}]{Allen08}
{Allen}, S.~W., {Rapetti}, D.~A., {Schmidt}, R.~W., {et~al.} 2008, \mnras, 383,
  879

\bibitem[{{Anders} \& {Grevesse}(1989)}]{1989GeCoA..53..197A}
{Anders}, E., \& {Grevesse}, N. 1989, \gca, 53, 197

\bibitem[{{Austermann} {et~al.}(2012){Austermann}, {Aird}, {Beall}, {Becker},
  {Bender}, {Benson}, {Bleem}, {Britton}, {Carlstrom}, {Chang}, {Chiang},
  {Cho}, {Crawford}, {Crites}, {Datesman}, {de Haan}, {Dobbs}, {George},
  {Halverson}, {Harrington}, {Henning}, {Hilton}, {Holder}, {Holzapfel},
  {Hoover}, {Huang}, {Hubmayr}, {Irwin}, {Keisler}, {Kennedy}, {Knox}, {Lee},
  {Leitch}, {Li}, {Lueker}, {Marrone}, {McMahon}, {Mehl}, {Meyer}, {Montroy},
  {Natoli}, {Nibarger}, {Niemack}, {Novosad}, {Padin}, {Pryke}, {Reichardt},
  {Ruhl}, {Saliwanchik}, {Sayre}, {Schaffer}, {Shirokoff}, {Stark}, {Story},
  {Vanderlinde}, {Vieira}, {Wang}, {Williamson}, {Yefremenko}, {Yoon}, \&
  {Zahn}}]{SPTPol12}
{Austermann}, J.~E., {Aird}, K.~A., {Beall}, J.~A., {et~al.} 2012, in
  \procspie, Vol. 8452, Millimeter, Submillimeter, and Far-Infrared Detectors
  and Instrumentation for Astronomy VI, 84521E

\bibitem[{{Avestruz} {et~al.}(2016){Avestruz}, {Nagai}, \& {Lau}}]{Avestruz16}
{Avestruz}, C., {Nagai}, D., \& {Lau}, E.~T. 2016, ArXiv e-prints

\bibitem[{{Balucinska-Church} \& {McCammon}(1992)}]{1992ApJ...400..699B}
{Balucinska-Church}, M., \& {McCammon}, D. 1992, \apj, 400, 699

\bibitem[{{Bartelmann} \& {Schneider}(2001)}]{Bartelmann01}
{Bartelmann}, M., \& {Schneider}, P. 2001, \physrep, 340, 291

\bibitem[{{Battaglia} {et~al.}(2013){Battaglia}, {Bond}, {Pfrommer}, \&
  {Sievers}}]{Battaglia13}
{Battaglia}, N., {Bond}, J.~R., {Pfrommer}, C., \& {Sievers}, J.~L. 2013, \apj,
  777, 123

\bibitem[{{Becker} \& {Kravtsov}(2011)}]{Becker11}
{Becker}, M.~R., \& {Kravtsov}, A.~V. 2011, \apj, 740, 25

\bibitem[{{Bleem} {et~al.}(2015){Bleem}, {Stalder}, {de Haan}, {Aird}, {Allen},
  {Applegate}, {Ashby}, {Bautz}, {Bayliss}, {Benson}, {Bocquet}, {Brodwin},
  {Carlstrom}, {Chang}, {Chiu}, {Cho}, {Clocchiatti}, {Crawford}, {Crites},
  {Desai}, {Dietrich}, {Dobbs}, {Foley}, {Forman}, {George}, {Gladders},
  {Gonzalez}, {Halverson}, {Hennig}, {Hoekstra}, {Holder}, {Holzapfel},
  {Hrubes}, {Jones}, {Keisler}, {Knox}, {Lee}, {Leitch}, {Liu}, {Lueker},
  {Luong-Van}, {Mantz}, {Marrone}, {McDonald}, {McMahon}, {Meyer}, {Mocanu},
  {Mohr}, {Murray}, {Padin}, {Pryke}, {Reichardt}, {Rest}, {Ruel}, {Ruhl},
  {Saliwanchik}, {Saro}, {Sayre}, {Schaffer}, {Schrabback}, {Shirokoff},
  {Song}, {Spieler}, {Stanford}, {Staniszewski}, {Stark}, {Story}, {Stubbs},
  {Vanderlinde}, {Vieira}, {Vikhlinin}, {Williamson}, {Zahn}, \&
  {Zenteno}}]{SPTSZ15}
{Bleem}, L.~E., {Stalder}, B., {de Haan}, T., {et~al.} 2015, \apjs, 216, 27

\bibitem[{{Broadhurst} {et~al.}(2005){Broadhurst}, {Takada}, {Umetsu}, {Kong},
  {Arimoto}, {Chiba}, \& {Futamase}}]{Broadhurst05}
{Broadhurst}, T., {Takada}, M., {Umetsu}, K., {et~al.} 2005, \apjl, 619, L143

\bibitem[{{Cappelluti} {et~al.}(2011){Cappelluti}, {Predehl}, {B{\"o}hringer},
  {Brunner}, {Brusa}, {Burwitz}, {Churazov}, {Dennerl}, {Finoguenov},
  {Freyberg}, {Friedrich}, {Hasinger}, {Kenziorra}, {Kreykenbohm}, {Lamer},
  {Meidinger}, {M{\"u}hlegger}, {Pavlinsky}, {Robrade}, {Santangelo},
  {Schmitt}, {Schwope}, {Steinmitz}, {Str{\"u}der}, {Sunyaev}, \&
  {Tenzer}}]{eROSITA11}
{Cappelluti}, N., {Predehl}, P., {B{\"o}hringer}, H., {et~al.} 2011, Memorie
  della Societa Astronomica Italiana Supplementi, 17, 159

\bibitem[{{Carrasco Kind} \& {Brunner}(2014)}]{MLZ14}
{Carrasco Kind}, M., \& {Brunner}, R.~J. 2014, \mnras, 438, 3409

\bibitem[{{Cavaliere} \& {Fusco-Femiano}(1976)}]{Cavaliere76}
{Cavaliere}, A., \& {Fusco-Femiano}, R. 1976, \aap, 49, 137

\bibitem[{{Chiu} {et~al.}(2016){Chiu}, {Mohr}, {McDonald}, {Bocquet}, {Ashby},
  {Bayliss}, {Benson}, {Bleem}, {Brodwin}, {Desai}, {Dietrich}, {Forman},
  {Gangkofner}, {Gonzalez}, {Hennig}, {Liu}, {Reichardt}, {Saro}, {Stalder},
  {Stanford}, {Song}, {Schrabback}, {{\v S}uhada}, {Strazzullo}, \&
  {Zenteno}}]{Chiu16}
{Chiu}, I., {Mohr}, J., {McDonald}, M., {et~al.} 2016, \mnras, 455, 258

\bibitem[{{Dark Energy Survey Collaboration} {et~al.}(2016){Dark Energy Survey
  Collaboration}, {Abbott}, {Abdalla}, {Aleksi{\'c}}, {Allam}, {Amara},
  {Bacon}, {Balbinot}, {Banerji}, {Bechtol}, {Benoit-L{\'e}vy}, {Bernstein},
  {Bertin}, {Blazek}, {Bonnett}, {Bridle}, {Brooks}, {Brunner}, {Buckley-Geer},
  {Burke}, {Caminha}, {Capozzi}, {Carlsen}, {Carnero-Rosell}, {Carollo},
  {Carrasco-Kind}, {Carretero}, {Castander}, {Clerkin}, {Collett}, {Conselice},
  {Crocce}, {Cunha}, {D'Andrea}, {da Costa}, {Davis}, {Desai}, {Diehl},
  {Dietrich}, {Dodelson}, {Doel}, {Drlica-Wagner}, {Estrada}, {Etherington},
  {Evrard}, {Fabbri}, {Finley}, {Flaugher}, {Foley}, {Fosalba}, {Frieman},
  {Garc{\'{\i}}a-Bellido}, {Gaztanaga}, {Gerdes}, {Giannantonio}, {Goldstein},
  {Gruen}, {Gruendl}, {Guarnieri}, {Gutierrez}, {Hartley}, {Honscheid}, {Jain},
  {James}, {Jeltema}, {Jouvel}, {Kessler}, {King}, {Kirk}, {Kron}, {Kuehn},
  {Kuropatkin}, {Lahav}, {Li}, {Lima}, {Lin}, {Maia}, {Makler}, {Manera},
  {Maraston}, {Marshall}, {Martini}, {McMahon}, {Melchior}, {Merson}, {Miller},
  {Miquel}, {Mohr}, {Morice-Atkinson}, {Naidoo}, {Neilsen}, {Nichol}, {Nord},
  {Ogando}, {Ostrovski}, {Palmese}, {Papadopoulos}, {Peiris}, {Peoples},
  {Percival}, {Plazas}, {Reed}, {Refregier}, {Romer}, {Roodman}, {Ross},
  {Rozo}, {Rykoff}, {Sadeh}, {Sako}, {S{\'a}nchez}, {Sanchez}, {Santiago},
  {Scarpine}, {Schubnell}, {Sevilla-Noarbe}, {Sheldon}, {Smith}, {Smith},
  {Soares-Santos}, {Sobreira}, {Soumagnac}, {Suchyta}, {Sullivan}, {Swanson},
  {Tarle}, {Thaler}, {Thomas}, {Thomas}, {Tucker}, {Vieira}, {Vikram},
  {Walker}, {Wechsler}, {Weller}, {Wester}, {Whiteway}, {Wilcox}, {Yanny},
  {Zhang}, \& {Zuntz}}]{DES16}
{Dark Energy Survey Collaboration}, {Abbott}, T., {Abdalla}, F.~B., {et~al.}
  2016, \mnras, 460, 1270

\bibitem[{{Diemer} \& {Kravtsov}(2014)}]{Diemer14}
{Diemer}, B., \& {Kravtsov}, A.~V. 2014, ArXiv e-prints

\bibitem[{{Donahue} {et~al.}(2014){Donahue}, {Voit}, {Mahdavi}, {Umetsu},
  {Ettori}, {Merten}, {Postman}, {Hoffer}, {Baldi}, {Coe}, {Czakon},
  {Bartelmann}, {Benitez}, {Bouwens}, {Bradley}, {Broadhurst}, {Ford},
  {Gastaldello}, {Grillo}, {Infante}, {Jouvel}, {Koekemoer}, {Kelson}, {Lahav},
  {Lemze}, {Medezinski}, {Melchior}, {Meneghetti}, {Molino}, {Moustakas},
  {Moustakas}, {Nonino}, {Rosati}, {Sayers}, {Seitz}, {Van der Wel}, {Zheng},
  \& {Zitrin}}]{Donahue14}
{Donahue}, M., {Voit}, G.~M., {Mahdavi}, A., {et~al.} 2014, \apj, 794, 136

\bibitem[{{Eckert} {et~al.}(2016){Eckert}, {Ettori}, {Coupon}, {Gastaldello},
  {Pierre}, {Melin}, {Le Brun}, {McCarthy}, {Adami}, {Chiappetti}, {Faccioli},
  {Giles}, {Lavoie}, {Lef{\`e}vre}, {Lieu}, {Mantz}, {Maughan}, {McGee},
  {Pacaud}, {Paltani}, {Sadibekova}, {Smith}, \& {Ziparo}}]{Eckert16}
{Eckert}, D., {Ettori}, S., {Coupon}, J., {et~al.} 2016, \aap, 592, A12

\bibitem[{{Evrard}(1997)}]{Evrard97}
{Evrard}, A.~E. 1997, \mnras, 292, 289

\bibitem[{{Fujita} {et~al.}(2013){Fujita}, {Ohira}, \& {Yamazaki}}]{Fujita13}
{Fujita}, Y., {Ohira}, Y., \& {Yamazaki}, R. 2013, \apjl, 767, L4

\bibitem[{{Giodini} {et~al.}(2009){Giodini}, {Pierini}, {Finoguenov}, {Pratt},
  {Boehringer}, {Leauthaud}, {Guzzo}, {Aussel}, {Bolzonella}, {Capak}, {Elvis},
  {Hasinger}, {Ilbert}, {Kartaltepe}, {Koekemoer}, {Lilly}, {Massey},
  {McCracken}, {Rhodes}, {Salvato}, {Sanders}, {Scoville}, {Sasaki}, {Smolcic},
  {Taniguchi}, {Thompson}, \& {COSMOS Collaboration}}]{Giodini09}
{Giodini}, S., {Pierini}, D., {Finoguenov}, A., {et~al.} 2009, \apj, 703, 982

\bibitem[{{Hasselfield} {et~al.}(2013){Hasselfield}, {Hilton}, {Marriage},
  {Addison}, {Barrientos}, {Battaglia}, {Battistelli}, {Bond}, {Crichton},
  {Das}, {Devlin}, {Dicker}, {Dunkley}, {D{\"u}nner}, {Fowler}, {Gralla},
  {Hajian}, {Halpern}, {Hincks}, {Hlozek}, {Hughes}, {Infante}, {Irwin},
  {Kosowsky}, {Marsden}, {Menanteau}, {Moodley}, {Niemack}, {Nolta}, {Page},
  {Partridge}, {Reese}, {Schmitt}, {Sehgal}, {Sherwin}, {Sievers}, {Sif{\'o}n},
  {Spergel}, {Staggs}, {Swetz}, {Switzer}, {Thornton}, {Trac}, \&
  {Wollack}}]{ACTSZ13}
{Hasselfield}, M., {Hilton}, M., {Marriage}, T.~A., {et~al.} 2013, \jcap, 7,
  008

\bibitem[{{Heymans} {et~al.}(2006){Heymans}, {Van Waerbeke}, {Bacon}, {Berge},
  {Bernstein}, {Bertin}, {Bridle}, {Brown}, {Clowe}, {Dahle}, {Erben}, {Gray},
  {Hetterscheidt}, {Hoekstra}, {Hudelot}, {Jarvis}, {Kuijken}, {Margoniner},
  {Massey}, {Mellier}, {Nakajima}, {Refregier}, {Rhodes}, {Schrabback}, \&
  {Wittman}}]{Heymans06}
{Heymans}, C., {Van Waerbeke}, L., {Bacon}, D., {et~al.} 2006, \mnras, 368,
  1323

\bibitem[{{Hinshaw} {et~al.}(2013){Hinshaw}, {Larson}, {Komatsu}, {Spergel},
  {Bennett}, {Dunkley}, {Nolta}, {Halpern}, {Hill}, {Odegard}, {Page}, {Smith},
  {Weiland}, {Gold}, {Jarosik}, {Kogut}, {Limon}, {Meyer}, {Tucker}, {Wollack},
  \& {Wright}}]{WMAP09}
{Hinshaw}, G., {Larson}, D., {Komatsu}, E., {et~al.} 2013, \apjs, 208, 19

\bibitem[{{Hirata} \& {Seljak}(2003)}]{Hirata03}
{Hirata}, C., \& {Seljak}, U. 2003, \mnras, 343, 459

\bibitem[{{Hoekstra}(2003)}]{Hoekstra03}
{Hoekstra}, H. 2003, \mnras, 339, 1155

\bibitem[{{Hoekstra} {et~al.}(2015){Hoekstra}, {Herbonnet}, {Muzzin}, {Babul},
  {Mahdavi}, {Viola}, \& {Cacciato}}]{Hoekstra15}
{Hoekstra}, H., {Herbonnet}, R., {Muzzin}, A., {et~al.} 2015, {The Canadian
  Cluster Comparison Project: detailed study of systematics and updated weak
  lensing masses}

\bibitem[{{Ilbert} {et~al.}(2013){Ilbert}, {McCracken}, {Le F{\`e}vre},
  {Capak}, {Dunlop}, {Karim}, {Renzini}, {Caputi}, {Boissier}, {Arnouts},
  {Aussel}, {Comparat}, {Guo}, {Hudelot}, {Kartaltepe}, {Kneib}, {Krogager},
  {Le Floc'h}, {Lilly}, {Mellier}, {Milvang-Jensen}, {Moutard}, {Onodera},
  {Richard}, {Salvato}, {Sanders}, {Scoville}, {Silverman}, {Taniguchi},
  {Tasca}, {Thomas}, {Toft}, {Tresse}, {Vergani}, {Wolk}, \& {Zirm}}]{Ilbert13}
{Ilbert}, O., {McCracken}, H.~J., {Le F{\`e}vre}, O., {et~al.} 2013, \aap, 556,
  A55

\bibitem[{{Jones} \& {Forman}(1984)}]{Jones84}
{Jones}, C., \& {Forman}, W. 1984, \apj, 276, 38

\bibitem[{{Kalberla} {et~al.}(2005){Kalberla}, {Burton}, {Hartmann}, {Arnal},
  {Bajaja}, {Morras}, \& {P{\"o}ppel}}]{2005A&A...440..775K}
{Kalberla}, P.~M.~W., {Burton}, W.~B., {Hartmann}, D., {et~al.} 2005, \aap,
  440, 775

\bibitem[{{Kawaharada} {et~al.}(2010){Kawaharada}, {Okabe}, {Umetsu},
  {Takizawa}, {Matsushita}, {Fukazawa}, {Hamana}, {Miyazaki}, {Nakazawa}, \&
  {Ohashi}}]{Kawaharada10}
{Kawaharada}, M., {Okabe}, N., {Umetsu}, K., {et~al.} 2010, \apj, 714, 423

\bibitem[{{Kravtsov} {et~al.}(2005){Kravtsov}, {Nagai}, \&
  {Vikhlinin}}]{Kravtsov05}
{Kravtsov}, A.~V., {Nagai}, D., \& {Vikhlinin}, A.~A. 2005, \apj, 625, 588

\bibitem[{{Lagan{\'a}} {et~al.}(2011){Lagan{\'a}}, {Zhang}, {Reiprich}, \&
  {Schneider}}]{Lagana11}
{Lagan{\'a}}, T.~F., {Zhang}, Y.-Y., {Reiprich}, T.~H., \& {Schneider}, P.
  2011, \apj, 743, 13

\bibitem[{{Lapi} {et~al.}(2010){Lapi}, {Fusco-Femiano}, \&
  {Cavaliere}}]{Lapi10}
{Lapi}, A., {Fusco-Femiano}, R., \& {Cavaliere}, A. 2010, \aap, 516, A34

\bibitem[{{Lieu} {et~al.}(2016){Lieu}, {Smith}, {Giles}, {Ziparo}, {Maughan},
  {D{\'e}mocl{\`e}s}, {Pacaud}, {Pierre}, {Adami}, {Bah{\'e}}, {Clerc},
  {Chiappetti}, {Eckert}, {Ettori}, {Lavoie}, {Le Fevre}, {McCarthy},
  {Kilbinger}, {Ponman}, {Sadibekova}, \& {Willis}}]{Lieu2016}
{Lieu}, M., {Smith}, G.~P., {Giles}, P.~A., {et~al.} 2016, \aap, 592, A4

\bibitem[{{Lin} {et~al.}(2003){Lin}, {Mohr}, \& {Stanford}}]{Lin03}
{Lin}, Y.-T., {Mohr}, J.~J., \& {Stanford}, S.~A. 2003, \apj, 591, 749

\bibitem[{{Lin} {et~al.}(2004){Lin}, {Mohr}, \& {Stanford}}]{Lin04}
---. 2004, \apj, 610, 745

\bibitem[{{Lin} {et~al.}(2012){Lin}, {Stanford}, {Eisenhardt}, {Vikhlinin},
  {Maughan}, \& {Kravtsov}}]{Lin12}
{Lin}, Y.-T., {Stanford}, S.~A., {Eisenhardt}, P.~R.~M., {et~al.} 2012, \apjl,
  745, L3

\bibitem[{{Louis} {et~al.}(2016){Louis}, {Grace}, {Hasselfield}, {Lungu},
  {Maurin}, {Addison}, {Ade}, {Aiola}, {Allison}, {Amiri}, {Angile},
  {Battaglia}, {Beall}, {de Bernardis}, {Bond}, {Britton}, {Calabrese}, {Cho},
  {Choi}, {Coughlin}, {Crichton}, {Crowley}, {Datta}, {Devlin}, {Dicker},
  {Dunkley}, {D{\"u}nner}, {Ferraro}, {Fox}, {Gallardo}, {Gralla}, {Halpern},
  {Henderson}, {Hill}, {Hilton}, {Hilton}, {Hincks}, {Hlozek}, {Ho}, {Huang},
  {Hubmayr}, {Huffenberger}, {Hughes}, {Infante}, {Irwin}, {Muya Kasanda},
  {Klein}, {Koopman}, {Kosowsky}, {Li}, {Madhavacheril}, {Marriage}, {McMahon},
  {Menanteau}, {Moodley}, {Munson}, {Naess}, {Nati}, {Newburgh}, {Nibarger},
  {Niemack}, {Nolta}, {Nu{\~n}ez}, {Page}, {Pappas}, {Partridge}, {Rojas},
  {Schaan}, {Schmitt}, {Sehgal}, {Sherwin}, {Sievers}, {Simon}, {Spergel},
  {Staggs}, {Switzer}, {Thornton}, {Trac}, {Treu}, {Tucker}, {Van Engelen},
  {Ward}, \& {Wollack}}]{ACTPOL16}
{Louis}, T., {Grace}, E., {Hasselfield}, M., {et~al.} 2016, ArXiv e-prints

\bibitem[{{Mahdavi} {et~al.}(2013){Mahdavi}, {Hoekstra}, {Babul}, {Bildfell},
  {Jeltema}, \& {Henry}}]{Mahdavi13}
{Mahdavi}, A., {Hoekstra}, H., {Babul}, A., {et~al.} 2013, \apj, 767, 116

\bibitem[{{Mandelbaum} {et~al.}(2006){Mandelbaum}, {Seljak}, {Kauffmann},
  {Hirata}, \& {Brinkmann}}]{Mandelbaum06}
{Mandelbaum}, R., {Seljak}, U., {Kauffmann}, G., {Hirata}, C.~M., \&
  {Brinkmann}, J. 2006, \mnras, 368, 715

\bibitem[{{Mandelbaum} {et~al.}(2005){Mandelbaum}, {Hirata}, {Seljak}, {Guzik},
  {Padmanabhan}, {Blake}, {Blanton}, {Lupton}, \& {Brinkmann}}]{Mandelbaum05}
{Mandelbaum}, R., {Hirata}, C.~M., {Seljak}, U., {et~al.} 2005, \mnras, 361,
  1287

\bibitem[{{Mandelbaum} {et~al.}(2014){Mandelbaum}, {Rowe}, {Bosch}, {Chang},
  {Courbin}, {Gill}, {Jarvis}, {Kannawadi}, {Kacprzak}, {Lackner}, {Leauthaud},
  {Miyatake}, {Nakajima}, {Rhodes}, {Simet}, {Zuntz}, {Armstrong}, {Bridle},
  {Coupon}, {Dietrich}, {Gentile}, {Heymans}, {Jurling}, {Kent}, {Kirkby},
  {Margala}, {Massey}, {Melchior}, {Peterson}, {Roodman}, \&
  {Schrabback}}]{Mandelbaum14}
{Mandelbaum}, R., {Rowe}, B., {Bosch}, J., {et~al.} 2014, \apjs, 212, 5

\bibitem[{{Mandelbaum} {et~al.}(2015){Mandelbaum}, {Rowe}, {Armstrong}, {Bard},
  {Bertin}, {Bosch}, {Boutigny}, {Courbin}, {Dawson}, {Donnarumma}, {Fenech
  Conti}, {Gavazzi}, {Gentile}, {Gill}, {Hogg}, {Huff}, {Jee}, {Kacprzak},
  {Kilbinger}, {Kuntzer}, {Lang}, {Luo}, {March}, {Marshall}, {Meyers},
  {Miller}, {Miyatake}, {Nakajima}, {Ngol{\'e} Mboula}, {Nurbaeva}, {Okura},
  {Paulin-Henriksson}, {Rhodes}, {Schneider}, {Shan}, {Sheldon}, {Simet},
  {Starck}, {Sureau}, {Tewes}, {Zarb Adami}, {Zhang}, \&
  {Zuntz}}]{Mandelbaum15}
{Mandelbaum}, R., {Rowe}, B., {Armstrong}, R., {et~al.} 2015, \mnras, 450, 2963

\bibitem[{{Mandelbaum} {et~al.}(2017{\natexlab{a}}){Mandelbaum}, {Miyatake},
  {Hamana}, {Oguri}, {Simet}, {Armstrong}, {Bosch}, {Murata}, {Lanusse},
  {Leauthaud}, {Coupon}, {More}, {Takada}, {Miyazaki}, {Speagle}, {Shirasaki},
  {Sif{\'o}n}, {Huang}, {Nishizawa}, {Medezinski}, {Okura}, {Okabe}, {Czakon},
  {Takahashi}, {Coulton}, {Hikage}, {Komiyama}, {Lupton}, {Strauss}, {Tanaka},
  \& {Utsumi}}]{HSCWL1styr}
{Mandelbaum}, R., {Miyatake}, H., {Hamana}, T., {et~al.} 2017{\natexlab{a}},
  ArXiv:1705.06745

\bibitem[{{Mandelbaum} {et~al.}(2017{\natexlab{b}}){Mandelbaum}, {Lanusse},
  {Leauthaud}, {Armstrong}, {Simet}, {Miyatake}, {Meyers}, {Bosch}, {Miyazaki},
  \& {Tanaka}}]{HSCGREAT03}
{Mandelbaum}, R., {Lanusse}, F., {Leauthaud}, A., {et~al.} 2017{\natexlab{b}},
  ArXiv e-prints

\bibitem[{{Mantz} {et~al.}(2016){Mantz}, {Allen}, {Morris}, \&
  {Schmidt}}]{Mantz16a}
{Mantz}, A.~B., {Allen}, S.~W., {Morris}, R.~G., \& {Schmidt}, R.~W. 2016,
  \mnras, 456, 4020

\bibitem[{{Martino} {et~al.}(2014){Martino}, {Mazzotta}, {Bourdin}, {Smith},
  {Bartalucci}, {Marrone}, {Finoguenov}, \& {Okabe}}]{Martino14}
{Martino}, R., {Mazzotta}, P., {Bourdin}, H., {et~al.} 2014, \mnras, 443, 2342

\bibitem[{{Massey} {et~al.}(2007){Massey}, {Heymans}, {Berg{\'e}}, {Bernstein},
  {Bridle}, {Clowe}, {Dahle}, {Ellis}, {Erben}, {Hetterscheidt}, {High},
  {Hirata}, {Hoekstra}, {Hudelot}, {Jarvis}, {Johnston}, {Kuijken},
  {Margoniner}, {Mandelbaum}, {Mellier}, {Nakajima}, {Paulin-Henriksson},
  {Peeples}, {Roat}, {Refregier}, {Rhodes}, {Schrabback}, {Schirmer}, {Seljak},
  {Semboloni}, \& {van Waerbeke}}]{Massey07}
{Massey}, R., {Heymans}, C., {Berg{\'e}}, J., {et~al.} 2007, \mnras, 376, 13

\bibitem[{{Mazzotta} {et~al.}(2004){Mazzotta}, {Rasia}, {Moscardini}, \&
  {Tormen}}]{Mazzotta04}
{Mazzotta}, P., {Rasia}, E., {Moscardini}, L., \& {Tormen}, G. 2004, \mnras,
  354, 10

\bibitem[{{Medezinski} {et~al.}(2010){Medezinski}, {Broadhurst}, {Umetsu},
  {Oguri}, {Rephaeli}, \& {Ben{\'{\i}}tez}}]{Medezinski10}
{Medezinski}, E., {Broadhurst}, T., {Umetsu}, K., {et~al.} 2010, \mnras, 405,
  257

\bibitem[{{Medezinski} {et~al.}(2015){Medezinski}, {Umetsu}, {Okabe}, {Nonino},
  {Molnar}, {Massey}, {Dupke}, \& {Merten}}]{Medezinski15}
{Medezinski}, E., {Umetsu}, K., {Okabe}, N., {et~al.} 2015, ArXiv e-prints

\bibitem[{{Medezinski} {et~al.}(2017){Medezinski}, {Oguri}, {Nishizawa},
  {Speagle}, {Miyatake}, {Umetsu}, {Leauthaud}, {Murata}, {Mandelbaum},
  {Sif{\'o}n}, {Strauss}, {Huang}, {Simet}, {Okabe}, {Tanaka}, \&
  {Komiyama}}]{Medezinski17}
{Medezinski}, E., {Oguri}, M., {Nishizawa}, A.~J., {et~al.} 2017, ArXiv
  1706.00427

\bibitem[{{Melchior} {et~al.}(2016){Melchior}, {Gruen}, {McClintock}, {Varga},
  {Sheldon}, {Rozo}, {Amara}, {Becker}, {Benson}, {Bermeo}, {Bridle},
  {Clampitt}, {Dietrich}, {Hartley}, {Hollowood}, {Jain}, {Jarvis}, {Jeltema},
  {Kacprzak}, {MacCrann}, {Rykoff}, {Saro}, {Suchyta}, {Troxel}, {Zuntz},
  {Bonnett}, {Plazas}, {Abbott}, {Abdalla}, {Annis}, {Benoit-L{\'e}vy},
  {Bernstein}, {Bertin}, {Brooks}, {Buckley-Geer}, {Carnero Rosell}, {Carrasco
  Kind}, {Carretero}, {Cunha}, {D'Andrea}, {da Costa}, {Desai}, {Eifler},
  {Flaugher}, {Fosalba}, {Garc{\'{\i}}a-Bellido}, {Gaztanaga}, {Gerdes},
  {Gruendl}, {Gschwend}, {Gutierrez}, {Honscheid}, {James}, {Kirk}, {Krause},
  {Kuehn}, {Kuropatkin}, {Lahav}, {Lima}, {Maia}, {March}, {Martini},
  {Menanteau}, {Miller}, {Miquel}, {Mohr}, {Nichol}, {Ogando}, {Romer},
  {Sanchez}, {Scarpine}, {Sevilla-Noarbe}, {Smith}, {Soares-Santos},
  {Sobreira}, {Swanson}, {Tarle}, {Thomas}, {Walker}, {Weller}, \&
  {Zhang}}]{Melchior16}
{Melchior}, P., {Gruen}, D., {McClintock}, T., {et~al.} 2016, ArXiv e-prints

\bibitem[{{Meneghetti} {et~al.}(2010){Meneghetti}, {Rasia}, {Merten},
  {Bellagamba}, {Ettori}, {Mazzotta}, {Dolag}, \& {Marri}}]{Meneghetti10}
{Meneghetti}, M., {Rasia}, E., {Merten}, J., {et~al.} 2010, \aap, 514, A93

\bibitem[{{Miyatake} {et~al.}(2013){Miyatake}, {Nishizawa}, {Takada},
  {Mandelbaum}, {Mineo}, {Aihara}, {Spergel}, {Bickerton}, {Bond}, {Gralla},
  {Hajian}, {Hilton}, {Hincks}, {Hughes}, {Infante}, {Lin}, {Lupton},
  {Marriage}, {Marsden}, {Menanteau}, {Miyazaki}, {Moodley}, {Niemack},
  {Oguri}, {Price}, {Reese}, {Sif{\'o}n}, {Wollack}, \& {Yasuda}}]{Miyatake13}
{Miyatake}, H., {Nishizawa}, A.~J., {Takada}, M., {et~al.} 2013, \mnras, 429,
  3627

\bibitem[{{Miyatake et al.}(in prep)}]{Miyatake17}
{Miyatake et al.} in prep, \pasj

\bibitem[{{Miyazaki} {et~al.}(2012){Miyazaki}, {Komiyama}, {Nakaya}, {Kamata},
  {Doi}, {Hamana}, {Karoji}, {Furusawa}, {Kawanomoto}, {Morokuma}, {Ishizuka},
  {Nariai}, {Tanaka}, {Uraguchi}, {Utsumi}, {Obuchi}, {Okura}, {Oguri},
  {Takata}, {Tomono}, {Kurakami}, {Namikawa}, {Usuda}, {Yamanoi}, {Terai},
  {Uekiyo}, {Yamada}, {Koike}, {Aihara}, {Fujimori}, {Mineo}, {Miyatake},
  {Yasuda}, {Nishizawa}, {Saito}, {Tanaka}, {Uchida}, {Katayama}, {Wang},
  {Chen}, {Lupton}, {Loomis}, {Bickerton}, {Price}, {Gunn}, {Suzuki},
  {Miyazaki}, {Muramatsu}, {Yamamoto}, {Endo}, {Ezaki}, {Itoh}, {Miwa},
  {Yokota}, {Matsuda}, {Ebinuma}, \& {Takeshi}}]{Miyazaki12}
{Miyazaki}, S., {Komiyama}, Y., {Nakaya}, H., {et~al.} 2012, in \procspie, Vol.
  8446, Ground-based and Airborne Instrumentation for Astronomy IV, 84460Z

\bibitem[{{Miyazaki} {et~al.}(2015){Miyazaki}, {Oguri}, {Hamana}, {Tanaka},
  {Miller}, {Utsumi}, {Komiyama}, {Furusawa}, {Sakurai}, {Kawanomoto},
  {Nakata}, {Uraguchi}, {Koike}, {Tomono}, {Lupton}, {Gunn}, {Karoji},
  {Aihara}, {Murayama}, \& {Takada}}]{Miyazaki15}
{Miyazaki}, S., {Oguri}, M., {Hamana}, T., {et~al.} 2015, \apj, 807, 22

\bibitem[{{Navarro} {et~al.}(1996){Navarro}, {Frenk}, \& {White}}]{NFW96}
{Navarro}, J.~F., {Frenk}, C.~S., \& {White}, S.~D.~M. 1996, \apj, 462, 563

\bibitem[{{Navarro} {et~al.}(1997){Navarro}, {Frenk}, \& {White}}]{NFW97}
---. 1997, \apj, 490, 493

\bibitem[{{Oguri}(2014)}]{Oguri14b}
{Oguri}, M. 2014, \mnras, 444, 147

\bibitem[{{Oguri} {et~al.}(2012){Oguri}, {Bayliss}, {Dahle}, {Sharon},
  {Gladders}, {Natarajan}, {Hennawi}, \& {Koester}}]{Oguri12}
{Oguri}, M., {Bayliss}, M.~B., {Dahle}, H., {et~al.} 2012, \mnras, 420, 3213

\bibitem[{{Oguri} {et~al.}(2010){Oguri}, {Takada}, {Okabe}, \&
  {Smith}}]{Oguri10b}
{Oguri}, M., {Takada}, M., {Okabe}, N., \& {Smith}, G.~P. 2010, \mnras, 405,
  2215

\bibitem[{{Oguri} {et~al.}(2005){Oguri}, {Takada}, {Umetsu}, \&
  {Broadhurst}}]{Oguri05}
{Oguri}, M., {Takada}, M., {Umetsu}, K., \& {Broadhurst}, T. 2005, \apj, 632,
  841

\bibitem[{{Oguri} {et~al.}(2017){Oguri}, {Lin}, {Lin}, {Nishizawa}, {More},
  {More}, {Hsieh}, {Medezinski}, {Miyatake}, {Jian}, {Lin}, {Takada}, {Okabe},
  {Speagle}, {Coupon}, {Leauthaud}, {Lupton}, {Miyazaki}, {Price}, {Tanaka},
  {Chiu}, {Komiyama}, {Okura}, {Tanaka}, \& {Usuda}}]{Oguri17}
{Oguri}, M., {Lin}, Y.-T., {Lin}, S.-C., {et~al.} 2017, ArXiv e-prints

\bibitem[{{Okabe} {et~al.}(2014{\natexlab{a}}){Okabe}, {Futamase}, {Kajisawa},
  \& {Kuroshima}}]{Okabe14a}
{Okabe}, N., {Futamase}, T., {Kajisawa}, M., \& {Kuroshima}, R.
  2014{\natexlab{a}}, \apj, 784, 90

\bibitem[{{Okabe} \& {Smith}(2016)}]{Okabe16b}
{Okabe}, N., \& {Smith}, G.~P. 2016, \mnras, 461, 3794

\bibitem[{{Okabe} {et~al.}(2013){Okabe}, {Smith}, {Umetsu}, {Takada}, \&
  {Futamase}}]{Okabe13}
{Okabe}, N., {Smith}, G.~P., {Umetsu}, K., {Takada}, M., \& {Futamase}, T.
  2013, \apjl, 769, L35

\bibitem[{{Okabe} {et~al.}(2010){Okabe}, {Takada}, {Umetsu}, {Futamase}, \&
  {Smith}}]{Okabe10b}
{Okabe}, N., {Takada}, M., {Umetsu}, K., {Futamase}, T., \& {Smith}, G.~P.
  2010, \pasj, 62, 811

\bibitem[{{Okabe} \& {Umetsu}(2008)}]{Okabe08}
{Okabe}, N., \& {Umetsu}, K. 2008, \pasj, 60, 345

\bibitem[{{Okabe} {et~al.}(2014{\natexlab{b}}){Okabe}, {Umetsu}, {Tamura},
  {Fujita}, {Takizawa}, {Zhang}, {Matsushita}, {Hamana}, {Fukazawa},
  {Futamase}, {Kawaharada}, {Miyazaki}, {Mochizuki}, {Nakazawa}, {Ohashi},
  {Ota}, {Sasaki}, {Sato}, \& {Tam}}]{Okabe14b}
{Okabe}, N., {Umetsu}, K., {Tamura}, T., {et~al.} 2014{\natexlab{b}}, \pasj,
  66, 99

\bibitem[{{Okabe} {et~al.}(2016){Okabe}, {Umetsu}, {Tamura}, {Fujita},
  {Takizawa}, {Matsushita}, {Fukazawa}, {Futamase}, {Kawaharada}, {Miyazaki},
  {Mochizuki}, {Nakazawa}, {Ohashi}, {Ota}, {Sasaki}, {Sato}, \&
  {Tam}}]{Okabe16}
---. 2016, \mnras, 456, 4475

\bibitem[{{Owers} {et~al.}(2013){Owers}, {Baldry}, {Bauer}, {Bland-Hawthorn},
  {Brown}, {Cluver}, {Colless}, {Driver}, {Edge}, {Hopkins}, {van Kampen},
  {Lara-Lopez}, {Liske}, {Loveday}, {Pimbblet}, {Ponman}, \&
  {Robotham}}]{Owers13}
{Owers}, M.~S., {Baldry}, I.~K., {Bauer}, A.~E., {et~al.} 2013, \apj, 772, 104

\bibitem[{{Pierre} {et~al.}(2016){Pierre}, {Pacaud}, {Adami}, {Alis},
  {Altieri}, {Baran}, {Benoist}, {Birkinshaw}, {Bongiorno}, {Bremer}, {Brusa},
  {Butler}, {Ciliegi}, {Chiappetti}, {Clerc}, {Corasaniti}, {Coupon}, {De
  Breuck}, {Democles}, {Desai}, {Delhaize}, {Devriendt}, {Dubois}, {Eckert},
  {Elyiv}, {Ettori}, {Evrard}, {Faccioli}, {Farahi}, {Ferrari}, {Finet},
  {Fotopoulou}, {Fourmanoit}, {Gandhi}, {Gastaldello}, {Gastaud},
  {Georgantopoulos}, {Giles}, {Guennou}, {Guglielmo}, {Horellou}, {Husband},
  {Huynh}, {Iovino}, {Kilbinger}, {Koulouridis}, {Lavoie}, {Le Brun}, {Le
  Fevre}, {Lidman}, {Lieu}, {Lin}, {Mantz}, {Maughan}, {Maurogordato},
  {McCarthy}, {McGee}, {Melin}, {Melnyk}, {Menanteau}, {Novak}, {Paltani},
  {Plionis}, {Poggianti}, {Pomarede}, {Pompei}, {Ponman}, {Ramos-Ceja},
  {Ranalli}, {Rapetti}, {Raychaudury}, {Reiprich}, {Rottgering}, {Rozo},
  {Rykoff}, {Sadibekova}, {Santos}, {Sauvageot}, {Schimd}, {Sereno}, {Smith},
  {Smol{\v c}i{\'c}}, {Snowden}, {Spergel}, {Stanford}, {Surdej}, {Valageas},
  {Valotti}, {Valtchanov}, {Vignali}, {Willis}, \& {Ziparo}}]{XXL16}
{Pierre}, M., {Pacaud}, F., {Adami}, C., {et~al.} 2016, \aap, 592, A1

\bibitem[{{Piffaretti} {et~al.}(2011){Piffaretti}, {Arnaud}, {Pratt},
  {Pointecouteau}, \& {Melin}}]{Piffaretti11}
{Piffaretti}, R., {Arnaud}, M., {Pratt}, G.~W., {Pointecouteau}, E., \&
  {Melin}, J.-B. 2011, \aap, 534, A109

\bibitem[{{Planck Collaboration}(2015)}]{Planck15Cosmology}
{Planck Collaboration}. 2015, ArXiv e-prints

\bibitem[{{Planck Collaboration} {et~al.}(2015){Planck Collaboration}, {Ade},
  {Aghanim}, {Arnaud}, {Ashdown}, {Aumont}, {Baccigalupi}, {Banday},
  {Barreiro}, {Bartlett}, \& et~al.}]{Planck15CluterCosmology}
{Planck Collaboration}, {Ade}, P.~A.~R., {Aghanim}, N., {et~al.} 2015, ArXiv
  e-prints

\bibitem[{{Planelles} {et~al.}(2013){Planelles}, {Borgani}, {Dolag}, {Ettori},
  {Fabjan}, {Murante}, \& {Tornatore}}]{Planelles13}
{Planelles}, S., {Borgani}, S., {Dolag}, K., {et~al.} 2013, \mnras, 431, 1487

\bibitem[{{Pratt} {et~al.}(2010){Pratt}, {Arnaud}, {Piffaretti},
  {B{\"o}hringer}, {Ponman}, {Croston}, {Voit}, {Borgani}, \&
  {Bower}}]{Pratt10}
{Pratt}, G.~W., {Arnaud}, M., {Piffaretti}, R., {et~al.} 2010, \aap, 511, A85

\bibitem[{{Reyes} {et~al.}(2012){Reyes}, {Mandelbaum}, {Gunn}, {Nakajima},
  {Seljak}, \& {Hirata}}]{Reyes12}
{Reyes}, R., {Mandelbaum}, R., {Gunn}, J.~E., {et~al.} 2012, \mnras, 425, 2610

\bibitem[{{Schneider} {et~al.}(1998){Schneider}, {van Waerbeke}, {Jain}, \&
  {Kruse}}]{Schneider98}
{Schneider}, P., {van Waerbeke}, L., {Jain}, B., \& {Kruse}, G. 1998, \mnras,
  296, 873

\bibitem[{{Shan} {et~al.}(2012){Shan}, {Kneib}, {Tao}, {Fan}, {Jauzac},
  {Limousin}, {Massey}, {Rhodes}, {Thanjavur}, \& {McCracken}}]{CFHTLS12}
{Shan}, H., {Kneib}, J.-P., {Tao}, C., {et~al.} 2012, \apj, 748, 56

\bibitem[{{Simet} {et~al.}(2017){Simet}, {McClintock}, {Mandelbaum}, {Rozo},
  {Rykoff}, {Sheldon}, \& {Wechsler}}]{Simet17}
{Simet}, M., {McClintock}, T., {Mandelbaum}, R., {et~al.} 2017, \mnras, 466,
  3103

\bibitem[{{Smith} {et~al.}(2016){Smith}, {Mazzotta}, {Okabe}, {Ziparo},
  {Mulroy}, {Babul}, {Finoguenov}, {McCarthy}, {Lieu}, {Bah{\'e}}, {Bourdin},
  {Evrard}, {Futamase}, {Haines}, {Jauzac}, {Marrone}, {Martino}, {May},
  {Taylor}, \& {Umetsu}}]{Smith16}
{Smith}, G.~P., {Mazzotta}, P., {Okabe}, N., {et~al.} 2016, \mnras, 456, L74

\bibitem[{{Smith} {et~al.}(2001){Smith}, {Brickhouse}, {Liedahl}, \&
  {Raymond}}]{2001ApJ...556L..91S}
{Smith}, R.~K., {Brickhouse}, N.~S., {Liedahl}, D.~A., \& {Raymond}, J.~C.
  2001, \apjl, 556, L91

\bibitem[{{Snowden} {et~al.}(2008){Snowden}, {Mushotzky}, {Kuntz}, \&
  {Davis}}]{Snowden2008}
{Snowden}, S.~L., {Mushotzky}, R.~F., {Kuntz}, K.~D., \& {Davis}, D.~S. 2008,
  \aap, 478, 615

\bibitem[{{Snowden} {et~al.}(1997){Snowden}, {Egger}, {Freyberg}, {McCammon},
  {Plucinsky}, {Sanders}, {Schmitt}, {Tr{\"u}mper}, \& {Voges}}]{RASS1997}
{Snowden}, S.~L., {Egger}, R., {Freyberg}, M.~J., {et~al.} 1997, \apj, 485, 125

\bibitem[{{Str{\"u}der} {et~al.}(2001){Str{\"u}der}, {Briel}, {Dennerl},
  {Hartmann}, {Kendziorra}, {Meidinger}, {Pfeffermann}, {Reppin}, {Aschenbach},
  {Bornemann}, {Br{\"a}uninger}, {Burkert}, {Elender}, {Freyberg}, {Haberl},
  {Hartner}, {Heuschmann}, {Hippmann}, {Kastelic}, {Kemmer}, {Kettenring},
  {Kink}, {Krause}, {M{\"u}ller}, {Oppitz}, {Pietsch}, {Popp}, {Predehl},
  {Read}, {Stephan}, {St{\"o}tter}, {Tr{\"u}mper}, {Holl}, {Kemmer}, {Soltau},
  {St{\"o}tter}, {Weber}, {Weichert}, {von Zanthier}, {Carathanassis}, {Lutz},
  {Richter}, {Solc}, {B{\"o}ttcher}, {Kuster}, {Staubert}, {Abbey}, {Holland},
  {Turner}, {Balasini}, {Bignami}, {La Palombara}, {Villa}, {Buttler},
  {Gianini}, {Lain{\'e}}, {Lumb}, \& {Dhez}}]{EPICPN}
{Str{\"u}der}, L., {Briel}, U., {Dennerl}, K., {et~al.} 2001, \aap, 365, L18

\bibitem[{{Sun} {et~al.}(2009){Sun}, {Voit}, {Donahue}, {Jones}, {Forman}, \&
  {Vikhlinin}}]{Sun09}
{Sun}, M., {Voit}, G.~M., {Donahue}, M., {et~al.} 2009, \apj, 693, 1142

\bibitem[{{Tanaka} {et~al.}(2017){Tanaka}, {Coupon}, {Hsieh}, {Mineo},
  {Nishizawa}, {Speagle}, {Furusawa}, {Miyazaki}, \& {Murayama}}]{HSCPhotoz17}
{Tanaka}, M., {Coupon}, J., {Hsieh}, B.-C., {et~al.} 2017, ArXiv e-prints

\bibitem[{{Tinker} {et~al.}(2010){Tinker}, {Robertson}, {Kravtsov}, {Klypin},
  {Warren}, {Yepes}, \& {Gottl{\"o}ber}}]{Tinker10}
{Tinker}, J.~L., {Robertson}, B.~E., {Kravtsov}, A.~V., {et~al.} 2010, \apj,
  724, 878

\bibitem[{{Turner} {et~al.}(2001){Turner}, {Abbey}, {Arnaud}, {Balasini},
  {Barbera}, {Belsole}, {Bennie}, {Bernard}, {Bignami}, {Boer}, {Briel},
  {Butler}, {Cara}, {Chabaud}, {Cole}, {Collura}, {Conte}, {Cros}, {Denby},
  {Dhez}, {Di Coco}, {Dowson}, {Ferrando}, {Ghizzardi}, {Gianotti}, {Goodall},
  {Gretton}, {Griffiths}, {Hainaut}, {Hochedez}, {Holland}, {Jourdain},
  {Kendziorra}, {Lagostina}, {Laine}, {La Palombara}, {Lortholary}, {Lumb},
  {Marty}, {Molendi}, {Pigot}, {Poindron}, {Pounds}, {Reeves}, {Reppin},
  {Rothenflug}, {Salvetat}, {Sauvageot}, {Schmitt}, {Sembay}, {Short},
  {Spragg}, {Stephen}, {Str{\"u}der}, {Tiengo}, {Trifoglio}, {Tr{\"u}mper},
  {Vercellone}, {Vigroux}, {Villa}, {Ward}, {Whitehead}, \& {Zonca}}]{EPICMOS1}
{Turner}, M.~J.~L., {Abbey}, A., {Arnaud}, M., {et~al.} 2001, \aap, 365, L27

\bibitem[{{Umetsu} {et~al.}(2011){Umetsu}, {Broadhurst}, {Zitrin},
  {Medezinski}, {Coe}, \& {Postman}}]{Umetsu11}
{Umetsu}, K., {Broadhurst}, T., {Zitrin}, A., {et~al.} 2011, \apj, 738, 41

\bibitem[{{Umetsu} {et~al.}(2016){Umetsu}, {Zitrin}, {Gruen}, {Merten},
  {Donahue}, \& {Postman}}]{Umetsu16}
{Umetsu}, K., {Zitrin}, A., {Gruen}, D., {et~al.} 2016, \apj, 821, 116

\bibitem[{{Vikhlinin} {et~al.}(2006){Vikhlinin}, {Kravtsov}, {Forman}, {Jones},
  {Markevitch}, {Murray}, \& {Van Speybroeck}}]{Vikhlinin06}
{Vikhlinin}, A., {Kravtsov}, A., {Forman}, W., {et~al.} 2006, \apj, 640, 691

\bibitem[{{Vikhlinin} {et~al.}(2009{\natexlab{a}}){Vikhlinin}, {Burenin},
  {Ebeling}, {Forman}, {Hornstrup}, {Jones}, {Kravtsov}, {Murray}, {Nagai},
  {Quintana}, \& {Voevodkin}}]{Vikhlinin09a}
{Vikhlinin}, A., {Burenin}, R.~A., {Ebeling}, H., {et~al.} 2009{\natexlab{a}},
  \apj, 692, 1033

\bibitem[{{Vikhlinin} {et~al.}(2009{\natexlab{b}}){Vikhlinin}, {Kravtsov},
  {Burenin}, {Ebeling}, {Forman}, {Hornstrup}, {Jones}, {Murray}, {Nagai},
  {Quintana}, \& {Voevodkin}}]{Vikhlinin09b}
{Vikhlinin}, A., {Kravtsov}, A.~V., {Burenin}, R.~A., {et~al.}
  2009{\natexlab{b}}, \apj, 692, 1060

\bibitem[{{von der Linden} {et~al.}(2014){von der Linden}, {Mantz}, {Allen},
  {Applegate}, {Kelly}, {Morris}, {Wright}, {Allen}, {Burchat}, {Burke},
  {Donovan}, \& {Ebeling}}]{vonderLinden14}
{von der Linden}, A., {Mantz}, A., {Allen}, S.~W., {et~al.} 2014, \mnras, 443,
  1973

\bibitem[{{Walker} {et~al.}(2012){Walker}, {Fabian}, {Sanders}, \&
  {George}}]{Walker12b}
{Walker}, S.~A., {Fabian}, A.~C., {Sanders}, J.~S., \& {George}, M.~R. 2012,
  \mnras, 427, L45

\bibitem[{{Yang} {et~al.}(2006){Yang}, {Mo}, {van den Bosch}, {Jing},
  {Weinmann}, \& {Meneghetti}}]{Yang06}
{Yang}, X., {Mo}, H.~J., {van den Bosch}, F.~C., {et~al.} 2006, \mnras, 373,
  1159

\bibitem[{{Zhang} {et~al.}(2008){Zhang}, {Finoguenov}, {B{\"o}hringer},
  {Kneib}, {Smith}, {Kneissl}, {Okabe}, \& {Dahle}}]{Zhang08}
{Zhang}, Y.-Y., {Finoguenov}, A., {B{\"o}hringer}, H., {et~al.} 2008, \aap,
  482, 451

\bibitem[{{Zhang} {et~al.}(2010){Zhang}, {Okabe}, {Finoguenov}, {Smith},
  {Piffaretti}, {Valdarnini}, {Babul}, {Evrard}, {Mazzotta}, {Sanderson}, \&
  {Marrone}}]{Zhang10}
{Zhang}, Y.-Y., {Okabe}, N., {Finoguenov}, A., {et~al.} 2010, \apj, 711, 1033

\end{thebibliography}

\appendix

\section{Results of Spectral fit} \label{sec:Appspecfit}

We summarize results of simultaneous fit for the spectrum in Table
\ref{table:temp}. The technical details are described in Sec. \ref{sec:xray}.

\begin{table*}
  \caption{Cluster details. $^a$Cluster name. $^b$Cluster-centric annulus $^c$
 counts in the energy band of 0.3-11 KeV of each instrument.
 $^{d}$ Best-fit temperature and $^{e}$ Best-fit abundance} \label{table:temp}
 \begin{tabular}{lclllcc}
      \hline
      \hline
  Name$^a$ & Annulus$^b$ &  \multicolumn{3}{c}{counts$\pm$error$^c$} & Temperature$^d$ & Abandance$^e$ \\
           & (arcsec) & MOS1 & MOS2 & PN & (keV) & ($Z_\odot$)\\
\hline
MCXCJ0157.4-0550
 & 0- 60 
 & 834%$\pm$30
 & 829%$\pm$30
 & 1400%$\pm$40
 & $ 3.51 _{- 0.26 }^{+ 0.32 }$
 & $ 0.26 $\\
 & 60-100  
 & 1038%$\pm$34
 & 1163%$\pm$37
 & 1975%$\pm$49
 & $ 3.02 _{- 0.22 }^{+ 0.27 }$
 & $ 0.18 $\\
 & 100-140
 & 1196%$\pm$38
 & 1267%$\pm$39
 & 1955%$\pm$50
 & $ 3.28 _{- 0.25 }^{+ 0.33 }$
 & $ 0.18 $\\
 & 140-180
 & 1228%$\pm$40
 & 1303%$\pm$41
 & 2085%$\pm$53
 & $ 2.84 _{- 0.23 }^{+ 0.30 }$
 & $ 0.20 $\\
 & 180-270
 & 2867%$\pm$64
 & 2732%$\pm$64
 & 4647%$\pm$85
 & $ 2.73 _{- 0.13 }^{+ 0.28 }$
 & $ 0.16 $\\
 & 270-360
 & 2920%$\pm$68
 & 2640%$\pm$66
 & 5306%$\pm$95
 & $ 2.54 _{- 0.10 }^{+ 0.31 }$
 & $ 0.10 $\\
 & 360-600
 & 7778%$\pm$115
 & 8959%$\pm$129
 & 17871%$\pm$178
 & -
 & \\
 & 600-900
 & 3117%$\pm$85
 & 5036%$\pm$121
 & 13795%$\pm$185
 & -
 & \\
MCXCJ0231.7-0451
 & 0- 40
 & 1411%$\pm$37
 & 1355%$\pm$37
 & 3084%$\pm$56
 & $ 5.64 _{- 0.34 }^{+ 0.40 }$
 & $ 0.52 $\\
 & 40- 60
 & 1381%$\pm$37
 & 1353%$\pm$37
 & 2739%$\pm$53
 &$ 5.03 _{- 0.37 }^{+ 0.37 }$
 &$ 0.12 $\\
 & 60- 80
 & 1339%$\pm$37
 & 1229%$\pm$35
 & 2492%$\pm$51
 &$ 4.33 _{- 0.27 }^{+ 0.36 }$
 &$ 0.46 $\\
 & 80-100
 & 1182%$\pm$35
 & 1198%$\pm$35
 & 2455%$\pm$51
 &$ 5.03 _{- 0.47 }^{+ 0.52 }$
 &$ 0.33 $\\
 & 100-140
 & 1936%$\pm$46
 & 1882%$\pm$46
 & 3737%$\pm$65
 &$ 4.09 _{- 0.22 }^{+ 0.23 }$
 &$ 0.20 $\\
 & 140-180
 & 1155%$\pm$37
 & 1174%$\pm$37
 & 2052%$\pm$51
 &$ 4.50 _{- 0.44 }^{+ 0.55 }$
 &$ 0.20 $\\
 & 180-270
 & 1665%$\pm$50
 & 1591%$\pm$50
 & 3373%$\pm$72
 &$ 3.10 _{- 0.38 }^{+ 0.54 }$
 &$ 0.20 $\\
 & 270-400
 & 1633%$\pm$55
 & 1798%$\pm$63
 & 4002%$\pm$90
 &-
 &\\
MCXCJ0201.7-0212
 & 0- 40
 & 7690%$\pm$87
 & 7369%$\pm$86
 & 16494%$\pm$128
 &$ 3.30 _{- 0.05 }^{+ 0.05 }$
 & $ 0.41 $\\
 & 40- 60 
 & 2617%$\pm$51
 & 2379%$\pm$49
 & 4139%$\pm$65
 &$ 4.23 _{- 0.27 }^{+ 0.36 }$
 &$ 0.28 $ \\
 & 60- 80 
 & 1649%$\pm$41
 & 1686%$\pm$42
 & 3218%$\pm$58
 &$ 4.26 _{- 0.36 }^{+ 0.48 }$
 &$ 0.14 $ \\
 & 80-100 
 & 1157%$\pm$35
 & 1151%$\pm$35
 & 2440%$\pm$52
 &$ 4.47 _{- 0.69 }^{+ 0.58 }$
 & $ 0.14 $\\
 & 100-140 
 & 1667%$\pm$44
 & 1710%$\pm$45
 & 3198%$\pm$62
 &$ 3.38 _{- 0.20 }^{+ 0.41 }$
 &$ 0.17 $ \\
 & 140-180 
 & 1044%$\pm$39
 & 1034%$\pm$38
 & 1750%$\pm$51
 &$ 4.28 _{- 0.49 }^{+ 0.67 }$
 & $ 0.17 $\\
 & 180-270 
 & 1596%$\pm$55
 & 1719%$\pm$56
 & 2486%$\pm$72
 &$ 2.43 _{- 0.28 }^{+ 0.27 }$
 &$ 0.17 $ \\
 & 270-360 
 & 1324%$\pm$57
 & 1502%$\pm$58
 & 2463%$\pm$79
 &-
 & \\
% & 360-540 
% & 1929$\pm$83
% & 2792$\pm$93
% & 6809$\pm$130
% &-
%& \\
MCXCJ1415.2-0030
 & 0- 50 
 & 509%$\pm$23
 & 519%$\pm$23
 & 889%$\pm$31
 &$ 3.12 _{- 0.25 }^{+ 0.25 }$
 &$ 0.22 $\\
 & 50- 90 
 & 690%$\pm$27
 & 683%$\pm$28
 & 1102%$\pm$36
 &$ 3.99 _{- 0.41 }^{+ 0.46 }$
 &$ 0.29 $\\
 & 90-140 
 & 761%$\pm$31
 & 736%$\pm$31
 & 1522%$\pm$44
 &$ 3.00 _{- 0.34 }^{+ 0.31 }$
 &$ 0.47 $\\
 & 140-180 
 & 481%$\pm$26
 & 444%$\pm$25
 & 890%$\pm$37
 &$ 2.03 _{- 0.38 }^{+ 0.54 }$
 &$ 0.21 $\\
 & 180-270 
 & 969%$\pm$40
 & 953%$\pm$41
 & 1984%$\pm$59
 &$ 1.71 _{- 0.13 }^{+ 0.12 }$
 &$ 0.12 $\\
 & 270-360 
 & 1302%$\pm$48
 & 1225%$\pm$48
 & 2653%$\pm$69
 &-
 &-\\
% & 360-540
% & 1691$\pm$61
% & 2679$\pm$78
% & 7672$\pm$117
% & -
% &\\
MCXCJ1415.2-0030W
 & 0-80
 & 191%$\pm$14
 & 229%$\pm$16
 & 417%$\pm$22
 &$2.06_{-0.22}^{+0.29}$
 &$ 0.34 $\\
 & 80-140
 & 198%$\pm$16
 & 246%$\pm$18
 & 430%$\pm$24
 &$1.80_{-0.28}^{+0.94}$
 &$ 0.33 $\\
 & 140-270
 & 772%$\pm$35
 & 803%$\pm$37
 & 1473%$\pm$49
 & -
  &   \\
  \hline
    \end{tabular}
\end{table*}

\end{document}